\documentclass[fleqn,usenatbib]{mnras}

\usepackage{newtxtext,newtxmath}

\usepackage[T1]{fontenc}

\DeclareRobustCommand{\VAN}[3]{#2}
\let\VANthebibliography\thebibliography
\def\thebibliography{\DeclareRobustCommand{\VAN}[3]{##3}\VANthebibliography}

\usepackage{graphicx}	
\usepackage{amsmath}	
\usepackage{stfloats}
\usepackage{multirow}
\usepackage[normalem]{ulem}
\usepackage{threeparttable}

\title[\textit{JWST} Mid-IR luminosity functions]{Exploring the faintest end of mid-infrared luminosity functions up to $z\simeq 5$ with the \textit{JWST} CEERS survey}

\author[Ling et al. 2024]{
Chih-Teng Ling$^{1}$,
Tomotsugu Goto$^{1,2}$,
Seong Jin Kim$^{1}$,
Cossas K.-W. Wu$^{1,2}$,
Tetsuya Hashimoto$^{3}$,
\newauthor
Tom C.-C. Chien$^{2}$,
Yu-Wei Lin$^{2}$,
Simon C.-C. Ho$^{4,5,6,7}$, and
Ece Kilerci$^{8}$
\\
$^{1}$Institute of Astronomy, National Tsing Hua University, 101, Section 2. Kuang-Fu Road, Hsinchu, 30013, Taiwan (R.O.C.)\\
$^{2}$Department of Physics, National Tsing Hua University, 101, Section 2. Kuang-Fu Road, Hsinchu, 30013, Taiwan (R.O.C.)\\
$^{3}$Department of Physics, National Chung Hsing University, 145, Xingda Road, Taichung, 40227, Taiwan (R.O.C.)\\
$^{4}$Research School of Astronomy and Astrophysics, The Australian National University, Canberra, ACT 2611, Australia\\
$^{5}$Centre for Astrophysics and Supercomputing, Swinburne University of Technology, P.O. Box 218, Hawthorn, VIC 3122, Australia\\
$^{6}$OzGrav: The Australian Research Council Centre of Excellence for Gravitational Wave Discovery, Hawthorn, VIC 3122, Australia\\
$^{7}$ASTRO3D: The Australian Research Council Centre of Excellence for All-sky Astrophysics in 3D, ACT 2611, Australia\\
$^{8}$Sabanc{\i} University, Faculty of Engineering and Natural Sciences, 34956, Istanbul, Turkey
}

\date{Accepted 2024 February 5. Received 2023 December 29; in original form 2023 May 31}

\pubyear{2024}

\begin{document}
\label{firstpage}
\pagerange{\pageref{firstpage}--\pageref{lastpage}}
\maketitle

\begin{abstract}
Mid-infrared (MIR) light from galaxies is sensitive to dust-obscured star-formation activities because it traces the characteristic emission of dust heated by young, massive stars. By constructing the MIR luminosity functions (LFs), we are able to quantify the overall dusty star formation history and the evolution of galaxies over cosmic time. 
In this work, we report the first rest-frame MIR LFs at 7.7, 10, 12.8, 15, 18, and 21 $\mu$m as well as the total IR LF from the James Webb Space Telescope (\textit{JWST}) Cosmic Evolution Early Release Science (CEERS) survey. We identify 506 galaxies at $z=0-5.1$ in the CEERS survey that also have optical photometry from the Hubble Space Telescope. With the unprecedented sensitivity of the \textit{JWST}, we probe the faintest end of the LFs at $z=0-1$ down to $L^* \sim 10^7 L_\odot$, $\sim 2$ orders of magnitude fainter than those from the previous generation of IR space telescopes. 
Our findings connect well with and continue the faint end of the MIR LFs from the deepest observations in past works. As a proxy of star formation history, we present the MIR-based luminosity density up to $z\simeq4.0$, marking the first probe of the early Universe by \textit{JWST} MIRI.
\end{abstract}




\section{Introduction}
The galaxy luminosity function (LF) is a statistic that profiles the number density of luminosities of galaxies in a specific volume, redshift range, and population. Serving as one of the fundamental quantities in extra-galactic astronomy, LF provides crucial and direct information for studying the evolution and properties of galaxies. 
The star-forming (SF) activities of galaxies play a significant role in shaping and evolving LF, as the star formation rate (SFR) and the ultra-violet (UV) / infrared (IR) luminosity are linked by well-known empirical relations \citep{Kennicutt1998ARA&A..36..189K, Madau1998ApJ...498..106M}. 

While UV light provides direct evidence of SF, IR-based LF is vital for a comprehensive understanding of galaxy evolution because it can trace the dust-obscured star formation that is opaque to UV. This makes the IR-based LF particularly crucial for studying the high-$z$ galaxy population, where the dust-obscured star formation is expected to be more significant \citep[e.g.,][]{Hopkins2001AJ....122..288H}. 
Specifically, we focus on the mid-infrared (MIR) wavelength due to its unique ability to detect polycyclic aromatic hydrocarbon (PAH) emission from 3 to 20 $\mu$m. The emission, along with its characteristic features at approximately 6.2, 7.7, and 8.6 $\mu$m, is mainly due to the heated dust by young stars. The MIR/PAH emissions dominate the spectral energy distribution (SED) of SF galaxies and have already been proven as a good indicator of total IR (TIR) luminosity $L_{\rm TIR}$ (which is estimated by integrating the galaxy SED from 8 to 1000 $\mu$m), e.g., \citealt{Caputi2007, 2011MNRAS.410..573G} for 8 $\mu$m, \citealt{2005ApJ...630...82P} for 12 $\mu$m, and SFR \citep{Wu2005ApJ...632L..79W, Calzetti2005ApJ...633..871C, Calzetti2007ApJ...666..870C}.

Studies on the IR LF can be dated back to the first-generation IR space telescopes \textit{IRAS} \citep{Neugebauer1984} and \textit{ISO} \citep{Kessler1996}, and continue with \textit{AKARI} \citep{Murakami2007}, \textit{Spitzer} \citep{Werner2004}, and \textit{Herschel} \citep{Pilbratt2010}.
These works \citep[e.g.,][]{Saunders1990, Rowan-Robinson1997, 2006MNRAS.370.1159B, Caputi2007, Goto2010, 2011MNRAS.410..573G, 2019PASJ...71...30G} have proven that IR LF is a powerful tool to unravel the evolution in dusty galaxies and have greatly improved our understanding of the cosmic star formation history (CSFH). The evolution of IR LFs has been extended to the whole $z=0-4$ range since \textit{Herschel} era \citep{2013MNRAS.432...23G, 2013A&A...553A.132M}, and is recently pushed to $z\simeq6$ \citep{Gruppioni2020A&A...643A...8G} with ALMA.
With the launch of the James Webb Space Telescope \citep[\textit{JWST},][]{Gardner2006, Kalirai2018}, we are now able to expand our understanding of IR sources to an unprecedented degree. For instance, \cite{Donnan2023MNRAS.518.6011D} and \cite{Harikane2023ApJS..265....5H} suggest UV LFs at $z=8-15$ and $z=8-17$, respectively, with \textit{JWST} Near-Infrared Camera (NIRCam). On the other hand, source counts studied at $7-21$ $\mu$m MIR wavelengths covered by the Mid-Infrared Instrument \citep[MIRI,][]{Rieke2015b} have demonstrated exceptional sensitivity of MIRI to below $\mu$Jy level \citep[e.g.,][]{2022MNRAS.517..853L, Wu2023MNRAS.523.5187W}. \cite{Kim2023} interpret the cosmic star-formation history (CSFH) and black hole accretion history (BHAH) by applying the backward evolution of local LFs \citep{Gruppioni2011} to their source counts \citep{2022MNRAS.517..853L, Wu2023MNRAS.523.5187W}.

This study seeks to further the efforts from the preceding works by constructing the first IR LF from \textit{JWST} MIRI observations. The precise MIR photometry, combined with NIR and optical data from Hubble Space Telescope (\textit{HST}) and large ground-based telescopes, can provide a complete picture of SED for faint and distant galaxies.
We target to illustrate the MIR LF as well as its evolution to the faintest and oldest end that has not been entirely explored previously.
This paper is organised as follows: 
We introduce the \textit{JWST} MIR data, completeness, and the multi-wavelength merged catalogue we produced in \S \ref{S:data}.
The procedure to derive the LF is explained in \S \ref{S:analysis}.
In \S \ref{S:res}, we present our LFs and discuss their evolution.
The conclusion is given in \S \ref{S:conc}.
We adopt the {\it Planck18} cosmology \citep{Planck2020A&A...641A...6P}, i.e., $\Lambda$ cold dark matter cosmology with ($\Omega_{m}$, $\Omega_{\Lambda}$, $\Omega_{b}$, $h)=(0.310, 0.689, 0.0490, 0.677)$ throughout the paper. 

\section{Data}\label{S:data}
\subsection{MIRI observations}\label{s:miri}
We utilise images from \textit{JWST} Cosmic Evolution Early Release Science \citep[CEERS;][]{2017jwst.prop.1345F} survey to construct a MIR source catalogue. The CEERS survey is one of the Early Release Science programs of \textit{JWST}, providing the first observations from NIRCam, Near-Infrared Spectrograph (NIRSpec), and MIRI in the Extended Groth Strip (EGS) legacy field.
This work focuses on the MIRI pointings in the observation that were observed using 6 continuous broad-band filters (F770W, F1000W, F1280W, F1500W, F1800W and F2100W), covering a wavelength range from 7.7 $\mu$m to 21.0 $\mu$m. Unlike the proposed observation \citep{2017jwst.prop.1345F}, only two pointings (observation ID: o001\_t021 and o002\_t022, hereafter "the first two") have been observed using all 6 filters. F770W and F2100W observations are missing in the other two (o012\_t026 and o015\_t028, hereafter "the last two") pointings, leaving 4 available filters for the pointings. 
To obtain a larger sample, we use all the four pointings in this work, regardless of the waveband coverage in 7.7 and 21 $\mu$m.
The total sky coverage for the 4 pointings is 31380.56 arcsec$^2$ ($\sim$ 8.7 arcmin$^2$), and 15694.27 arcsec$^2$ for F770W and F2100W which have only 2 pointings.

We obtain level-3 image product of the pointings from the Mikulski Archive for Space Telescopes (MAST) and conduct source extraction and photometry following \cite{Wu2023MNRAS.523.5187W}. Two photometry software, \textsc{Photutils} \citep{Photutils_larry_bradley_2022_6825092} and \textsc{Source-Extractor} \citep{Bertin1996} are used in the procedure. \cite{Wu2023MNRAS.523.5187W} found that applying two software for background estimation and photometry separately is more effective. 
Please refer to \cite{Wu2023MNRAS.523.5187W} for the technical details of the photometry, including the effectiveness and the result compared to the public source catalogue from MAST. In order to keep a sufficient sample size, we do not select sources by a specific band. Instead, any sources with at least one detection in MIRI bands will be included in the catalogue. A total of 1210 sources are identified through this selection.

\subsection{Completeness}
Completeness correction is essential to source statistics. The completeness assesses the reliability of the source within an image, which is affected by the sensitivity and exposure time of the image.
\cite{Wu2023MNRAS.523.5187W} has presented an effective approach to probe the completeness of \textit{JWST} images.
Here, we measure the completeness function for all 4 CEERS pointings by performing Monte Carlo (MC) simulations with the same method as in \cite{Wu2023MNRAS.523.5187W} and \cite{Takagi2012}.
The method is briefly described as follows.
First, we randomly implement 20 artificial MIRI PSF sources with the same flux into the target image for each flux bin.
The width of the flux bin is $\Delta\log$ ($f_\nu$/Jy) 0.1 dex. The same photometric analysis is subsequently applied to these mock sources to determine their recovery rate as the final completeness.
The simulation was repeated 500 times to reduce the uncertainty.

\begin{figure}
    \centering
    \includegraphics[width=\columnwidth]{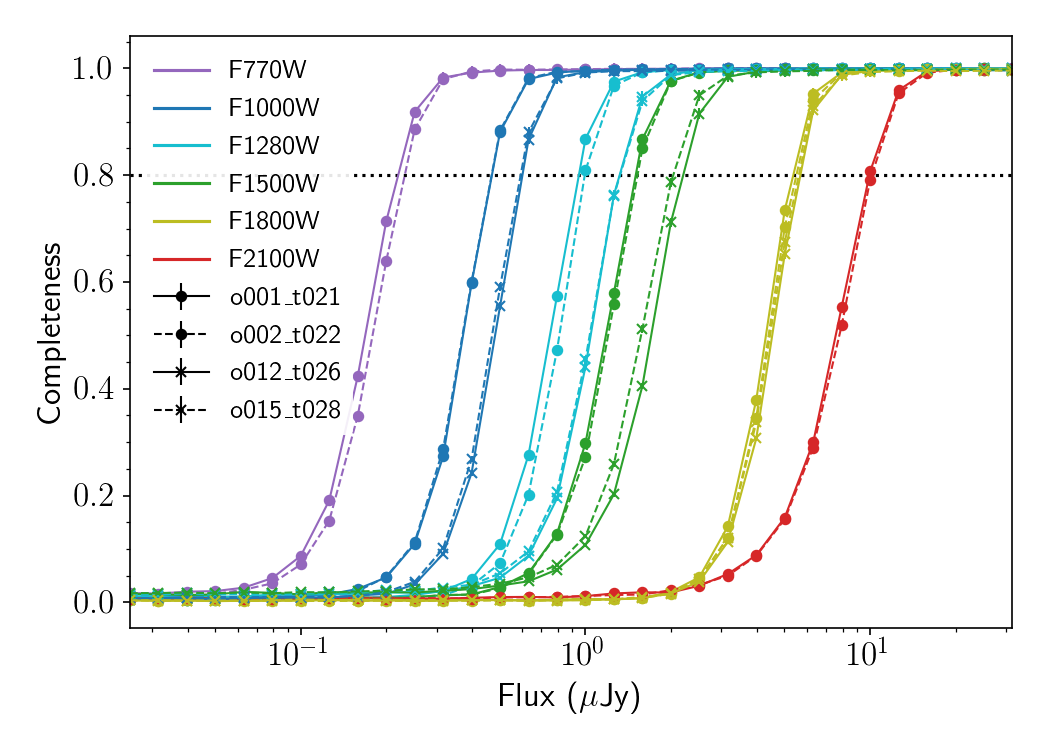}
    \caption{The measured completeness functions of the CEERS pointing o001\_t021, o002\_t022 (marked in circle), o012\_t026, o015\_t028 (marked in cross) images.
    The completeness of different bands is plotted in violet (F770W), blue (F1000W), cyan (F1280W), green (F1500W) yellow (F1800W), and red (F2100W), respectively. The horizontal black dotted line shows the 80\% completeness limit.}
    \label{fig:completeness}
\end{figure}

\begin{figure}
    \centering
    \includegraphics[width=\columnwidth]{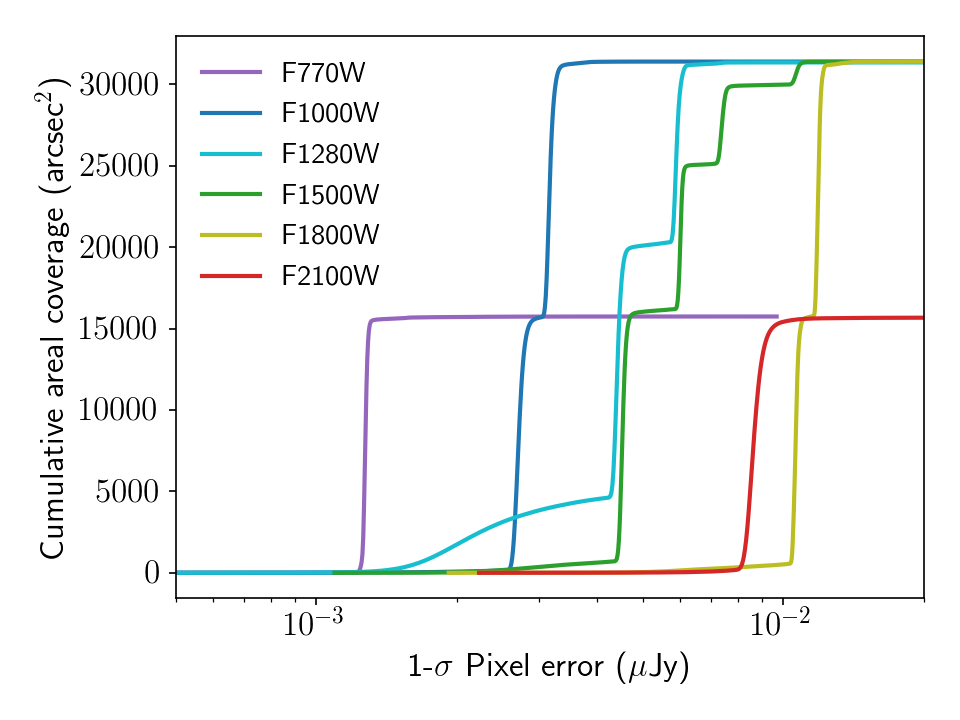}
    \caption{The cumulative areal coverage of the sum of all 4 pointings as a function of 1-sigma pixel error for each band. }
    \label{fig:areal}
\end{figure}

We build the completeness as a function of flux for each pointing and filter. Figure \ref{fig:completeness} shows completeness functions of the 4 CEERS pointings. 
The 80\% completeness limits for each pointing and MIRI band are concluded in Table \ref{tab:limit}.
These limits are taken by a linear interpolation between our data points in Figure \ref{fig:completeness}.
We find that the completeness and the 80\% limit of the last two pointings are generally higher than the first two.
This is due to the shorter exposure time in the last two pointings. The average exposure time of the last two pointings is 1243 seconds (F1000W and F1800W) and 932 seconds (F1280W and F1500W), while for the first two is 1673 seconds (except for F2100W, which has a much longer exposure time of 4811 seconds).
In Figure \ref{fig:areal}, we show the cumulative areal coverage of images of the sum of the 4 pointings as a function of 1 sigma flux error of each pixel, obtained from the fits images.
The vertically rising line in Figure \ref{fig:areal} indicates an identical pixel error for most individual pointings, thus ensuring the uniform exposure time. Meanwhile, the stepwise increase illustrates the difference in exposure time between the first two and last two pointings.
For F1280W and F1500W bands, we notice a more gradual rise because of variations in exposures within the last two pointings. 
We should note that no additional correction in completeness is applied to account for these variations because the random distribution of artificial sources in the MC simulation already includes and averages for such effects.

\begin{table}
    \centering
    \begin{tabular}{|c|c|c|c|c|}
        \hline
        & \multicolumn{2}{c}{\textbf{First two pointings}} & \multicolumn{2}{c}{\textbf{Last two pointings}} \\
        \textbf{Band} & \textbf{o001\_t021} & \textbf{o002\_t022} & \textbf{o012\_t026} & \textbf{o015\_t028} \\ \hline
        F770W & 0.22 $\mu$Jy & 0.23 $\mu$Jy & - & - \\
        F1000W & 0.47 $\mu$Jy & 0.47 $\mu$Jy & 0.60 $\mu$Jy & 0.59 $\mu$Jy \\
        F1280W & 0.95 $\mu$Jy & 0.99 $\mu$Jy & 1.3 $\mu$Jy & 1.3 $\mu$Jy \\
        F1500W & 1.5 $\mu$Jy & 1.5 $\mu$Jy & 2.2 $\mu$Jy & 2.0 $\mu$Jy \\
        F1800W & 5.4 $\mu$Jy & 5.5 $\mu$Jy & 5.7 $\mu$Jy & 5.6 $\mu$Jy \\
        F2100W & 10 $\mu$Jy & 10 $\mu$Jy & - & - \\
        \hline
    \end{tabular}
\caption{The 80\% completeness limit for each pointing and band.}
\label{tab:limit}
\end{table}

We inspect false/spurious detections in the images by evaluating their reliability, following \cite{Wu2023MNRAS.523.5187W}. 
The reliability is assessed as (N$_{\rm pos}$ - N$_{\rm neg}$) / N$_{\rm pos}$, where N$_{\rm pos}$ is the number of sources detected in the original images and N$_{\rm neg}$ is the number in the negative images.
The same photometry procedures are applied to the negative images (i.e., the original MIRI pointing multiplied by -1) to extract sources.
We show the reliability for each pointing and band beyond their 80\% completeness levels (refer to Table \ref{tab:limit}) in Table \ref{tab:spurious}.
All the pointing have a high level of reliability $>90\%$, except for F1280W and F1500W band in pointing o015\_t028. This is expected because of the non-uniform and shorter exposures in these pointings, as suggested above.

The completeness functions and reliability are accounted for in the derivation of LF to correct possible incompleteness and spurious detection of our sources. We elaborate on the correction procedure in \S \ref{S:lf}.

\begin{table}
    \centering
    \begin{tabular}{|c|c|c|c|c|}
        \hline
        & \multicolumn{2}{c}{\textbf{First two pointings}} & \multicolumn{2}{c}{\textbf{Last two pointings}} \\
        \textbf{Band} & \textbf{o001\_t021} & \textbf{o002\_t022} & \textbf{o012\_t026} & \textbf{o015\_t028} \\ \hline
        F770W & 93\% & 95\% & - & - \\
        F1000W & 97\% & 97\% & 95\% & 95\% \\
        F1280W & 96\% & 98\% & 92\% & 85\% \\
        F1500W & 90\% & 95\% & 93\% & 77\% \\
        F1800W & 100\% & 100\% & 100\% & 97\% \\
        F2100W & 90\% & 96\% & - & - \\
        \hline
    \end{tabular}
\caption{The reliability beyond the 80\% completeness limits for each pointing and band.}
\label{tab:spurious}
\end{table}

\subsection{Multi-wavelength merged catalogue}
A comprehensive photometric catalogue covering multiple wavelengths is necessary to accurately model the SEDs of our MIR sources. This is achieved by matching our source catalogue with the CANDELS-EGS Multi-wavelength catalogue \citep{2017ApJS..229...32S}. Built from the Cosmic Assembly Near-infrared Deep Extragalactic Legacy Survey \citep[CANDELS;][]{2011ApJS..197...35G, 2011ApJS..197...36K}, the CANDELS-EGS catalogue provides broad coverage of wavelengths from near-UV to MIR (0.4 $\mu$m $-$ 8 $\mu$m) in the EGS field. 
The CANDELS-EGS sources are selected by detections in \textit{HST} WFC3 F160W with a depth of 26.62 AB (90\% completeness).
An accompanying redshift catalogue \citep{2023ApJ...942...36K} of CANDELS-EGS sources is also combined for analysis in \S \ref{S:sed}. We apply a matching radius of 0.5 arcsec, considering the average size of our sources is $\sim 0.25$ arcsec. 94\% of identified pairs have a separation of less than 0.25 arcsec.

The compiled catalogue contains 573 sources with $15+6$ band photometry from the CANDELS-EGS catalogue and our \textit{JWST} MIR source catalogue. Table \ref{tab:1} summarises the depths in the catalogue. We should emphasise that the first two MIRI pointings do not include NIRCam photometry due to the lack of NIRCam observations in the pointings. Instead, we take the data from the CANDELS-EGS catalogue for NIR wavelengths, which are obtained by CFHT WIRCam, \textit{HST} WFC3, and \textit{Spitzer} IRAC (refer to Table \ref{tab:1}).

\begin{figure*}
    \centering
    \includegraphics[width=2\columnwidth]{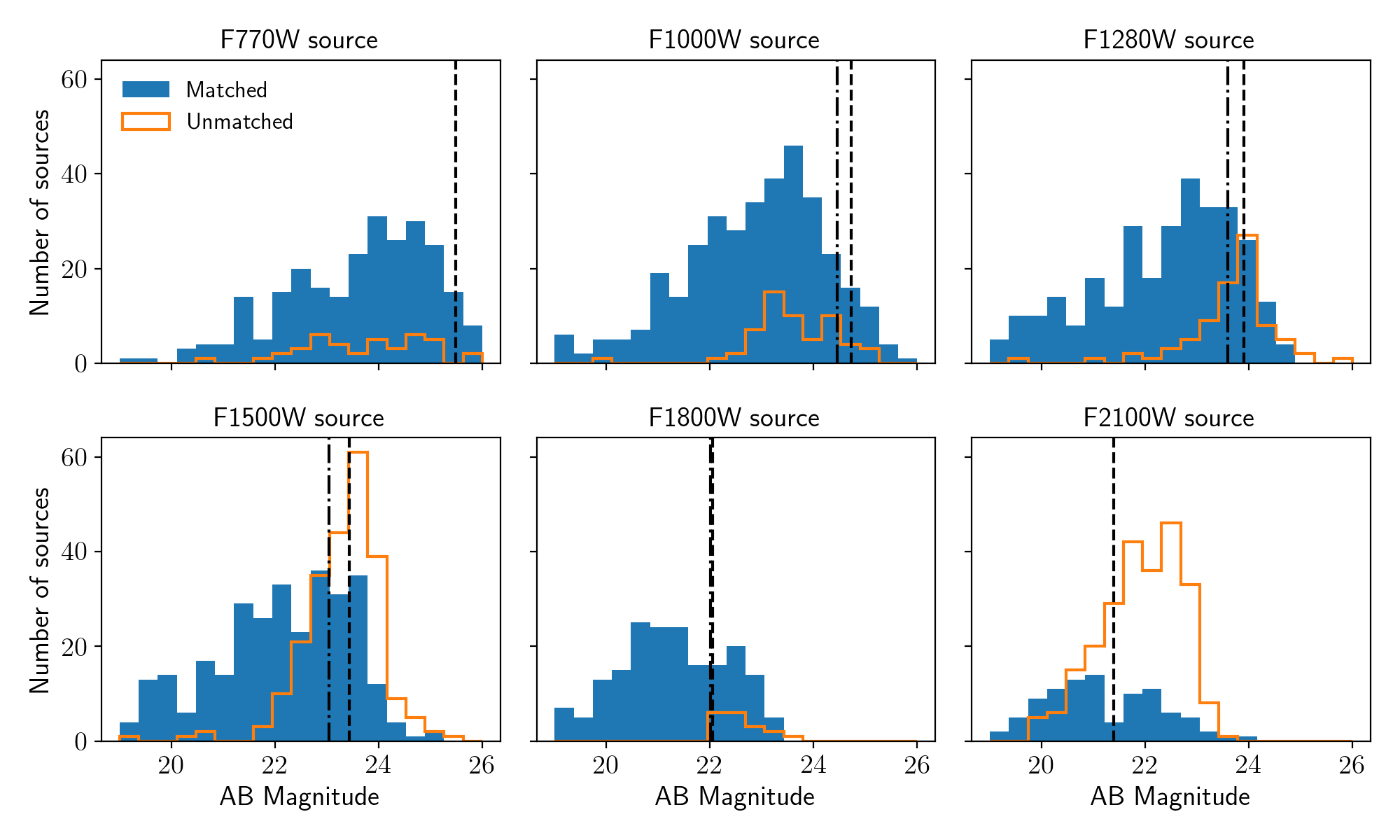}
    \caption{The magnitude histogram of all MIR source detection in each filter. Sources that are also identified in the CANDELS-EGS catalogue are marked as "match" (blue), otherwise, they are marked as "unmatched" (orange).
    The black dashed (dash-dotted) line indicates the 80\% completeness of the filter for pointing o002\_t022 (o015\_t028), converted to AB magnitude. Note that sources may be counted more than once here, as they are detected in multiple filters.}
    \label{fig:mag_mir_hist}
\end{figure*}

637 (52\% of the 1210 sources) MIR sources have not been identified in the CANDELS-EGS catalogue.
We exclude these sources from the compiled catalogue because their absence in optical/NIR bands could result in poor SED fit.
As shown in Figure \ref{fig:mag_mir_hist}, the majority of these sources (which come from F1280W, F1500W, and F2100W filters) mostly lie outside the 80\% completeness limit of the filters. 
For further check, we have eye-balled those EGS-undetected sources and found that most of them in F2100W are spurious among the noisy background, justifying the exclusion. 
This is also noted by the CEERS team \citep{Yang-MIRI-quality-2023ApJ...956L..12Y}, which shows the measured depth for F2100W is shallower than the Exposure Time Calculator prediction by $\sim 0.9$ magnitude.
In F1500W, about half of the EGS-undetected sources are close to the edge of the field, where the dithering of CANDELS-EGS and other MIRI filters just missed. 
Here we forewarn readers for the treatment of these spurious sources, especially in F2100W filter.

\begin{table*}
    \centering
    \begin{tabular}{ccccc}
        \hline
        \textbf{Telescope} & \textbf{Instrument} & \textbf{Bands} & \textbf{5$\sigma$ depth (AB)} & \textbf{Reference}\\ \hline
        CFHT & MegaCam & $u$* & 27.1 & \cite{2012AJ....143...38G} \\
        ...  & ... & $g^{\prime}$ & 27.3 & ... \\
        ...  & ... & $r^{\prime}$ & 27.2 & ... \\
        ...  & ... & $i^{\prime}$ & 27.0 & ... \\
        ...  & ... & $z^{\prime}$ & 26.1 & ... \\
        ...  & WIRCam & J & 24.4 & \cite{2012AA...545A..23B} \\
        ...  & ... & H & 24.5 & ... \\
        ...  & ... & K & 24.3 & ... \\
        \textit{HST} & ACS & F606W & 28.8 & \cite{2011ApJS..197...36K} \\
        ...  & ... & F814W & 28.2 & ... \\
        ...  & WFC3 & F125W & 27.6 & \cite{2011ApJS..197...36K} \\
        ...  & ... & F140W & 26.8 & ... \\
        ...  & ... & F160W & 27.6 & \cite{2014ApJS..214...24S} \\
        \textit{Spitzer} & IRAC & 3.6 $\mu$m & 23.9 & \cite{2015ApJS..218...33A} \\ 
        ...  & ... & 4.5 $\mu$m & 24.2 & ... \\ \hline
        \textit{JWST} & MIRI & F770W & 25.6 (25.5) & This work, \cite{Yang-MIRI-quality-2023ApJ...956L..12Y} \\ 
        ...  & ... & F1000W & 24.8 (24.7) & ... \\
        ...  & ... & F1280W & 24.1 (23.9) & ... \\
        ...  & ... & F1500W & 23.8 (23.4) & ... \\
        ...  & ... & F1800W & 22.8 (22.1) & ... \\
        ...  & ... & F2100W & 22.4 (21.4) & ... \\ \hline
    \end{tabular}
\caption{Depths in each filter in the compiled catalogue. The 5$\sigma$ depths of the CANDELS-EGS catalogue band are taken from \protect\cite{2017ApJS..229...32S}. For \textit{JWST} MIRI, we show the 5$\sigma$ depths of pointing o002\_t022 estimated by \protect\cite{Yang-MIRI-quality-2023ApJ...956L..12Y}. Our 80\% completeness limits for pointing o002\_t022 are presented in brackets as reference.}
\label{tab:1}
\end{table*}

\section{Analysis}\label{S:analysis}
There are several procedures involved in obtaining LF. We describe them step by step in the following subsections.
In \S \ref{S:sed}, we introduce the SED fitting results of our galaxies and evaluate their performance. In \S \ref{S:k-corr}, we present the conversion from the observed flux in the SED to the rest-frame luminosity with $K$-correction.
After the conversion, we show how we construct the rest-frame LFs in \S \ref{S:lf}, where the correction to the completeness of our galaxy sample is also addressed.

\subsection{SED fitting}\label{S:sed}
We utilise Code Investigating GALaxy Emission v2022.1 \citep[\textsc{cigale;}][]{2019A&A...622A.103B} to better estimate the photometric redshifts (photo-$z$) and SEDs for sources in the compiled catalogue by incorporating \textit{JWST} mid-IR photometry. 
\textsc{cigale} is a Python code that can model SEDs and physical properties of galaxies with observations across from far-UV to radio spectrum. \textsc{cigale} offers users various modules and parameters to optimise the fit. For our study, we mainly follow the fitting configuration described in \cite{Yang2023ApJ...950L...5Y}, which uses the same sample as us and analyses SEDs to investigate the Active Galactic Nuclei (AGN) population at high-$z$. The modules used for the fitting are presented in Table \ref{tab:CIGALE}. Unlisted parameters remain the default value from \textsc{cigale}.

The main difference between our parameters and \cite{Yang2023ApJ...950L...5Y} is that we adopt \texttt{dustatt\_modified\_CF00} module for dust attenuation, which reduces the systematically lower photo-$z$ from \textsc{cigale} \citep[Figure 2 in][]{Yang2023ApJ...950L...5Y}. We also utilise a denser redshift grid that ranges from 0.01 to 6.0 with 600 linearly spaced steps, yielding a step of 0.01 in redshift. This is preferred as the luminosity scales with the square of the distance, and a denser grid can reduce potential biases.

\begin{table*}
    \centering
     \begin{tabular}{lll}
        \hline
        \begin{tabular}[c]{@{}l@{}}\textbf{Module}\end{tabular} &
          \textbf{Parameters} &
          \textbf{Values} \\\hline
         &
          Stellar e-folding time {[}Gyr{]} &
          0.5, 1, 2, 5 \\
        \multirow{-2}{*}{\begin{tabular}[c]{@{}l@{}}Star formation history\\ \texttt{sfhdelayed}\end{tabular}} &
          Stellar age {[}Gyr{]} &
          1, 2, 5 \\\hline
         &
          Initial mass function &
          \cite{Salpeter1955} \\
        \multirow{-2}{*}{\begin{tabular}[c]{@{}l@{}}Simple Stellar population\\ \texttt{bc03}\end{tabular}} &
          Metallicity &
          0.02 \\\hline
         &
          Ionisation parameter {[}log{]} &
          $-2.0$ \\
        \multirow{-2}{*}{\begin{tabular}[c]{@{}l@{}}Nebular emission\\ \texttt{nebular}\end{tabular}} &
          Gas metallicity &
          0.2 \\\hline
         &
          V-band attenuation in the interstellar medium ($A_V^{\rm ISM}$) &
          0.01, 0.02, 0.04, 0.08, 0.16, 0.32, 0.63, 1.3, 2.5, 5, 10 \\
         &
          $A_V^{\rm ISM}$ / ($A_V^{\rm BC}+A_V^{\rm ISM}$) &
          0.44 \\
         &
          Power law slope of the attenuation in the ISM &
          $-0.9$, $-0.7$, $-0.5$ \\
        \multirow{-4}{*}{\begin{tabular}[c]{@{}l@{}}Dust attenuation\\ \texttt{dustatt\_modified\_CF00}\end{tabular}} &
          Power law slope of the attenuation in the birth clouds &
          $-1.3$, $-1.0$, $-0.7$ \\\hline
         &
          PAH mass fraction &
          0.47, 2.5, 7.32 \\
         &
          Minimum radiation field &
          0.1, 1.0, 10, 50 \\
        \multirow{-3}{*}{\begin{tabular}[c]{@{}l@{}}Galactic dust emission\\ \texttt{dl2014}\end{tabular}} &
          Fraction of PDR emission &
          0.01, 0.02, 0.05, 0.1, 0.2, 0.5, 0.9 \\\hline
         &
          Average edge-on optical depth at 9.7 $\mu$m &
          3, 5, 7, 9, 11 \\
         &
          Viewing angle &
          $70^{\circ}$ \\
         &
          AGN contribution to IR luminosity &
          0, 0.01, 0.03, 0.05, 0.1, 0.2, 0.3, 0.5, 0.75, 0.9, 0.99 \\
        \multirow{-4}{*}{\begin{tabular}[c]{@{}l@{}}AGN (UV-to-IR) emission\\ \texttt{skirtor2016}\end{tabular}} &
          Wavelength range where frac$_{\rm AGN}$ is defined &
          $3-30$ $\mu$m \\\hline
        \begin{tabular}[c]{@{}l@{}}Redshift+IGM\\ \texttt{redshift}\end{tabular} &
          redshift &
          $0.01-6.0$ (600 steps)
          \\\hline
        \end{tabular}
\caption{The modules and parameters used in \textsc{cigale}.}
\label{tab:CIGALE}
\end{table*}

The fitted galaxy SEDs are separated into SF and AGN types based on \textsc{cigale} parameter frac$_{\rm AGN}$.
frac$_{\rm AGN}$ is the AGN contribution to IR luminosity ratio in Table \ref{tab:CIGALE}, defined by $\frac{L_{\rm AGN}}{L_{\rm AGN} + L_{\rm galaxy}}$ within $3-30$ $\mu$m. 
Galaxies with frac$_{\rm AGN}$ exceeding 20\% are categorised as AGN hosts \citep[following][in which they used catalogues from \citealt{Kim2021MNRAS.500.4078K} and \citealt{Ho2021MNRAS.502..140H}]{Tina2020MNRAS.499.4068W}, otherwise as an SF galaxy. 133 (26\%) of our sample are identified as AGN host galaxies.
Examples of best-fit SED results from \textsc{cigale} for each redshift bin and SED type are presented in Figures \ref{fig:sed_example_00}-\ref{fig:sed_example_03}. Additionally, the fitted far-IR SED is included in these figures for reference. 

It is crucial to note that our far-IR SED fitting is dependent on observations at shorter MIR wavelengths due to the absence of far-IR detection; most galaxies are too faint to be observed with \textit{Herschel}. 
Despite the potential impact this may have in deriving $L_{\rm TIR}$, we would like to point out that $L_{\rm TIR}$ is constrained to a certain degree by $L_{\rm MIR}$\footnote{Here we use $L_{\rm MIR}$ in a broad sense as a luminosity in the mid-IR wavelength range, because the definition slightly differs depending on the literature.} or $L_{\rm PAH}$, especially for the SF galaxy, which is the majority of our sample. The empirical relation between $L_{\rm MIR}$ and $L_{\rm TIR}$ has been reported in the literature \citep[e.g.,][]{Caputi2007, Houck2007ApJ...671..323H, 2011MNRAS.410..573G, Lin2023}, where $L_{\rm MIR}$ has been demonstrated as a reliable proxy for $L_{\rm TIR}$ given that both luminosities are primarily attributed to SF activity. On this consideration, the derivation of $L_{\rm TIR}$ should remain valid and in line with the current understanding. 

We remind readers, however, that the assumption can only hold if both $L_{\rm MIR}$ and $L_{\rm TIR}$ are attributed to SF activity. If the AGN dominates the mid-IR emission, the contamination from AGN in the mid-IR can be as high as $80-100\%$ (e.g., the right panel of Figure \ref{fig:sed_example_03}). Therefore, in the presence of an AGN, the AGN component must be subtracted from the total SED fit, as we demonstrate in \S \ref{s:TIR}.

\begin{figure}
    \centering
    \includegraphics[width=\columnwidth]{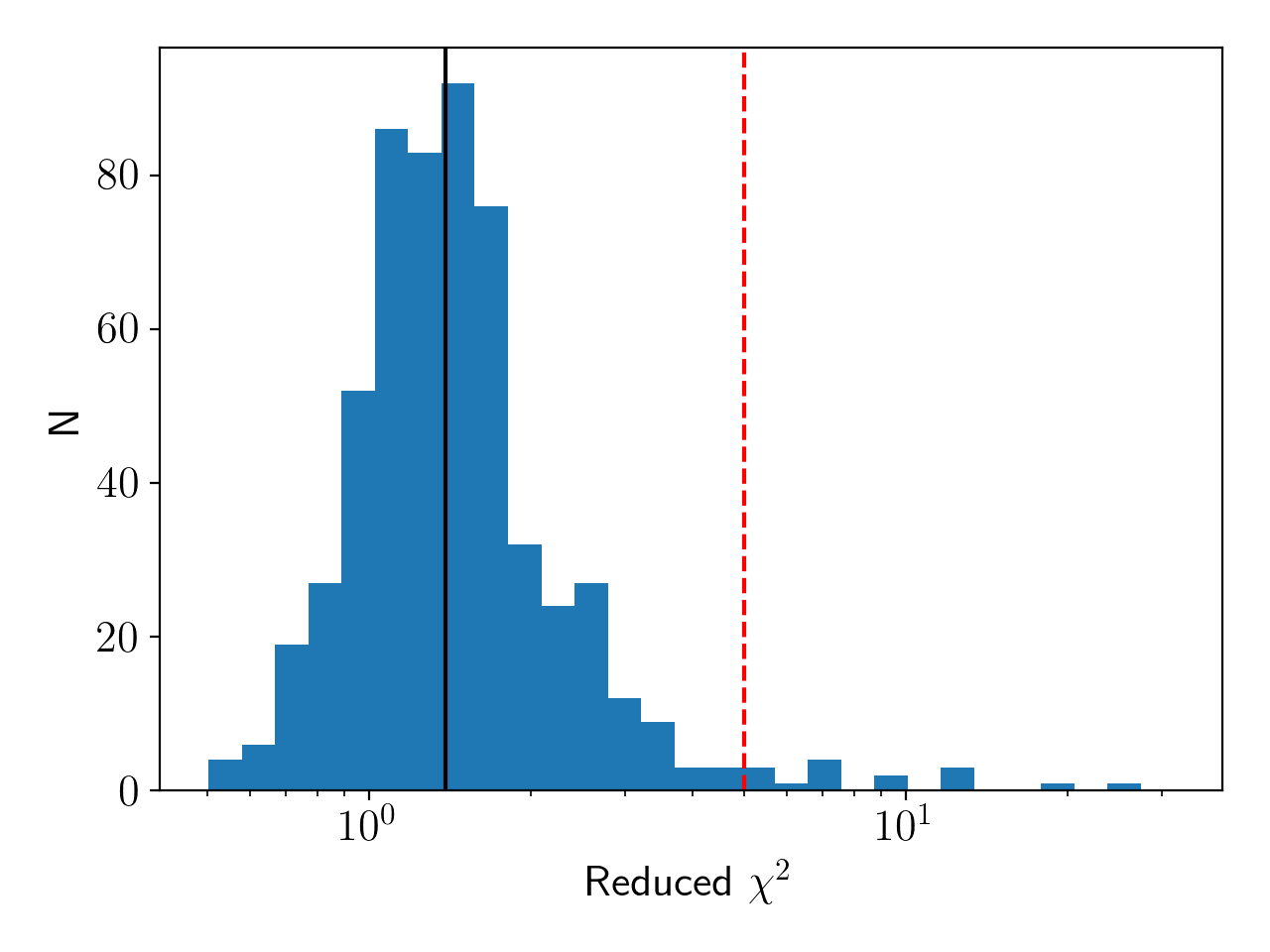}
    \caption{The reduced $\chi^2$ histogram of the fitting result. The black line is the median (1.39), and the red dashed line shows the criterion of $\chi^2=5$.}
    \label{fig:chi2_hist}
\end{figure}

\begin{figure}
    \centering
    \includegraphics[width=\columnwidth]{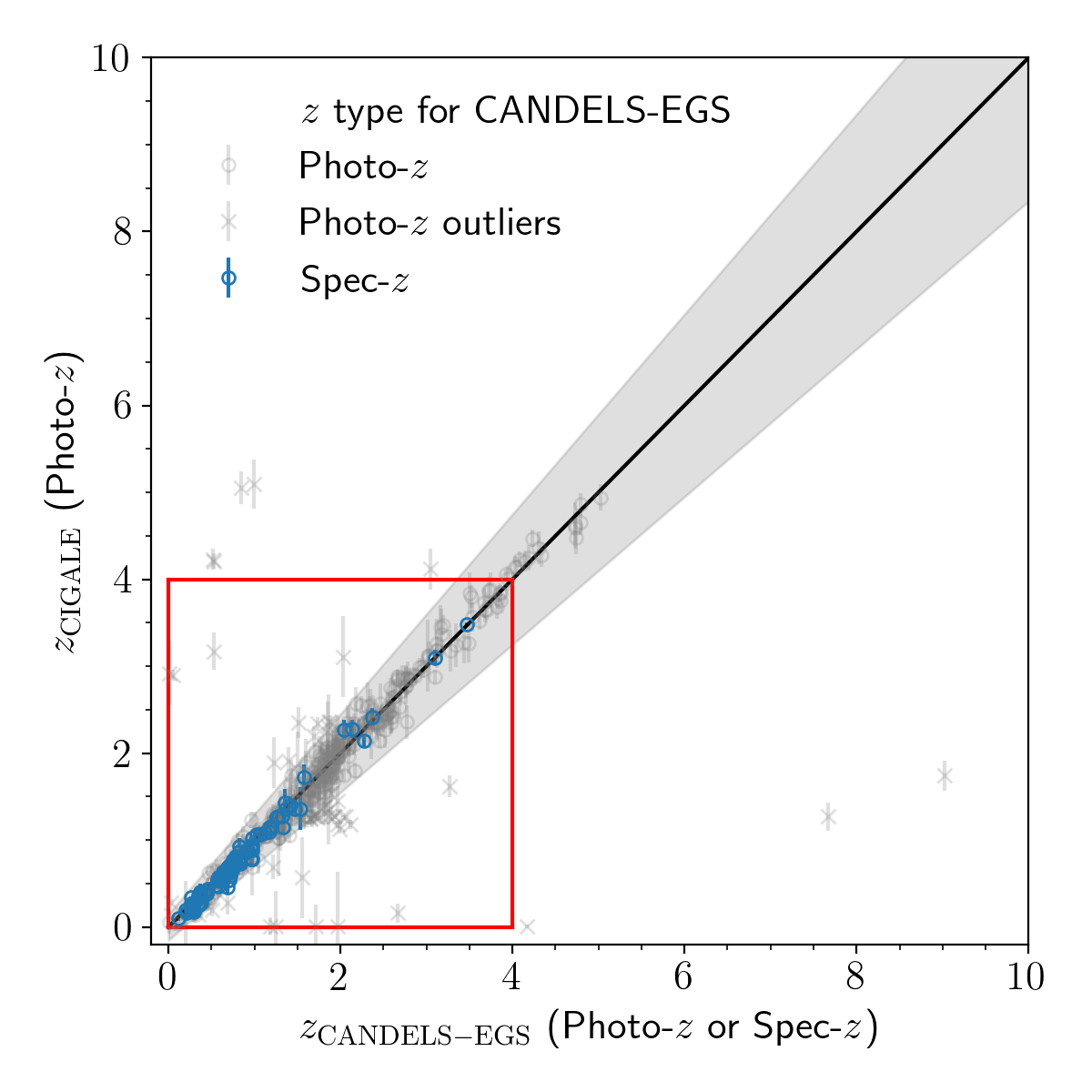}
    \includegraphics[width=\columnwidth]{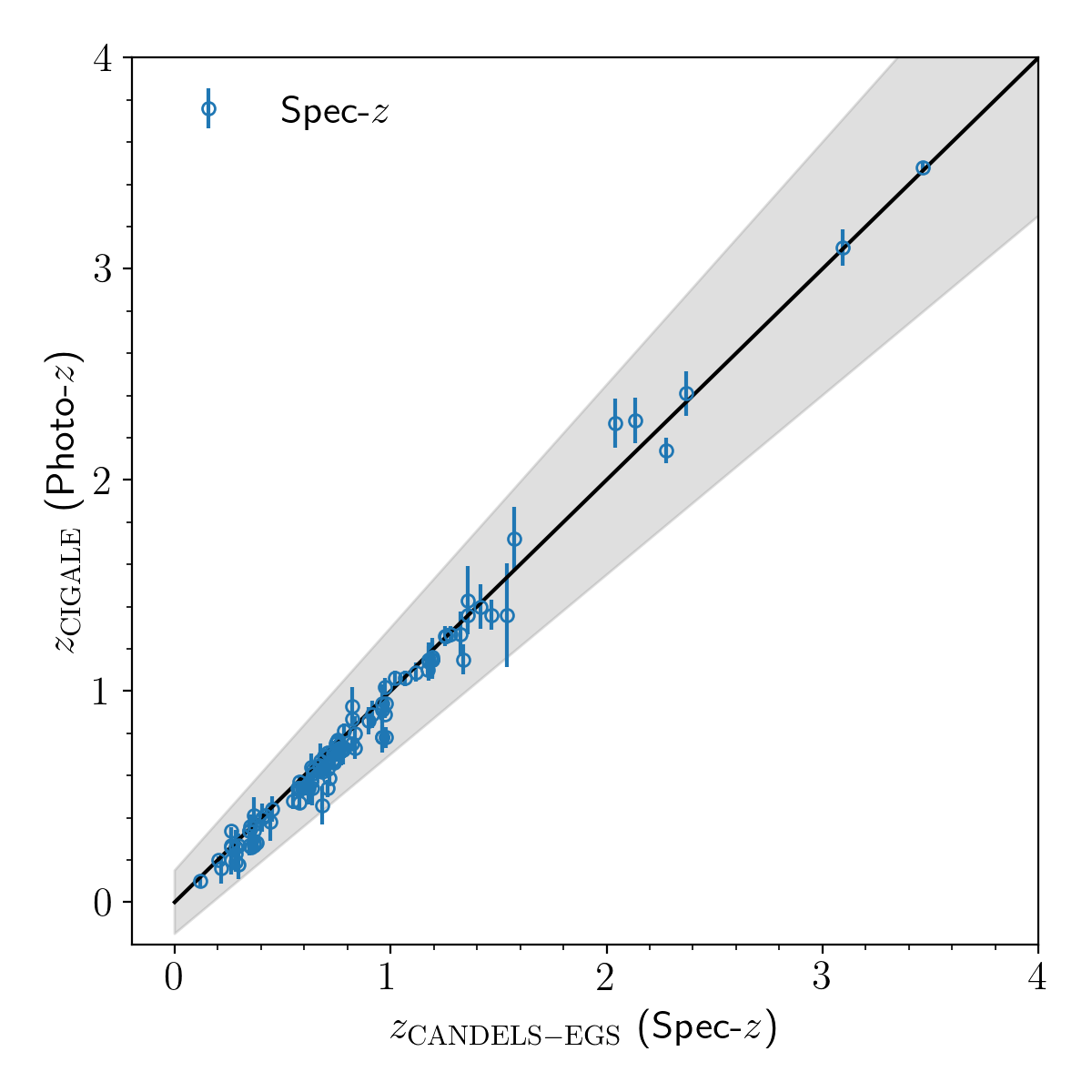}
    \caption{Estimated photo-$z$ from \textsc{cigale} ($z_{\rm CIGALE}$) versus redshift from the CANDELS-EGS catalogue ($z_{\rm CANDELS-EGS}$), where $z_{\rm CANDELS-EGS}$ can be either spec-$z$ or photo-$z$. The photo-$z$ value from the catalogue is used if spec-$z$ is unavailable. The two redshift types are presented separately in the upper panel, and outliers for photo-$z$ are shown in grey crosses. 1-to-1 line (black line) and 15\% uncertainty (grey region) are also plotted.
    The lower panel shows only sources with spec-$z$ available in the catalogue and is magnified to the proper range (indicated by the red square in the upper panel) for clarity.}
    \label{fig:z_compare}
\end{figure}

To examine the overall quality of the fit, we plot the distribution of reduced $\chi^2$ of each source's fit in Figure \ref{fig:chi2_hist}. The median of the distribution is 1.39. A criterion of reduced $\chi^2 < 5$ is then set to exclude 16 poor fits. We also present an analysis similar to \cite{Yang2023ApJ...950L...5Y} for redshift. In Figure \ref{fig:z_compare}, we compare the photo-$z$ estimated by \textsc{cigale} with the redshift catalogue \citep{2023ApJ...942...36K} which provides spectroscopic redshift (spec-$z$) measurements for 107 sources in our sample. The assessment of photo-$z$, i.e., the normalised median absolute deviation $\sigma_{\rm NMAD}$ (defined as $1.48 \times \, {\rm median} \{ |\Delta z - {\rm median}(\Delta z) | / (1+z_{\rm spec}) \}$) and spec-$z$ outlier fraction $\eta$ (defined as $\Delta z/(1+z_{\rm spec})>15\%$), can be derived by comparing these 107 galaxies with spec-$z$ (blue open circles in Figure \ref{fig:z_compare}) to their photo-$z$, $\Delta z = |z_{\rm photo} - z_{\rm spec}|$.
In result, we obtain $\sigma_{\rm NMAD} = 0.029$ and $\eta = 0.00\%$. 
These values are comparable to those reported in \cite{2023ApJ...942...36K} ($\sigma_{\rm NMAD} = 0.0227$; $\eta = 6.7\%$) and \cite{Yang2023ApJ...950L...5Y} ($\sigma_{\rm NMAD} = 0.031$; $\eta = 0.00\%$). 
We note, the 0\% outlier fraction \citep[as in][]{Yang2023ApJ...950L...5Y} simply reflects the fact that all the spec-$z$ samples have a photo-$z$ (from \textsc{cigale}) with $<15\%$ deviation.

For consistency, we excluded 51 (9.16\%) galaxies that are outside the 15\% redshift uncertainty range from our sample, i.e., the photo-$z$ outliers shown in the upper panel of Figure \ref{fig:z_compare}. 
In summary, we apply below criteria for the reduction:
\begin{itemize}
    \item Reduced $\chi^2$ of SED fitting $>5$
    \item Photo-$z$ outliers with $>15\%$ deviation from \cite{2023ApJ...942...36K}
\end{itemize}
This leaves 506 available photo-$z$ and SEDs for the following procedures. To compensate for the removal by the criteria, We multiply the luminosity functions in all the $z$-bins by a factor of  $573/506\simeq1.13$ in the following derivation (see \S \ref{S:lf}), with the underlying assumption that their redshift distribution is the same as the rest.

\subsection{$K$-correction}\label{S:k-corr}
Due to the expansion of the Universe, the observed fluxes are redshifted from the rest/emitted-frame. The difference in transmission efficiency between the two filters at different redshifts has to be considered to obtain accurate measurements for luminosity, especially since we are interested in the MIR spectra related to star formation. 

We apply the $K$-correction \citep{1968ApJ...154...21O} to correct the effect, with the correction factor $K(z)$ in terms of luminosity
 \begin{equation}
    K(z) = \frac{1}{1+z}\frac{\int d\lambda_{\rm o} \;\lambda_{\rm o} L_\lambda\left(\frac{\lambda_{\rm o}}{1+z}\right)S(\lambda_{\rm o})}{\int d\lambda_{\rm e} \;\lambda_{\rm e} L_\lambda\left(\lambda_{\rm e}\right)S(\lambda_{\rm e})}
\end{equation}
for the AB magnitude system. $\lambda_{\rm o}$ and $\lambda_{\rm e}$ are the filter wavelength in the observed and rest/emitted-frame respectively, and $S\left(\lambda\right)$ is the transmission curve for the specific filter. $L_\lambda$ is the luminosity density per unit wavelength (e.g., $L_\odot$ $\mu$m$^{-1}$).
The observed flux $F_\lambda$ and the rest-frame luminosity $L_\lambda$ are related by
\begin{align}
L_\lambda(\lambda_{\rm e}) &= \frac{4\pi D_{\rm L}(z)^2}{1+z} F_\lambda(\lambda_{\rm o}) \\
\lambda_{\rm e} &= \frac{\lambda_{\rm o}}{1+z}
\end{align}
where $D_{\rm L}$ is the luminosity distance at redshift $z$. 
For each filter, we convolve our best-fit galaxy SED in the rest-frame to the transmission curve of the filter to obtain the corrected luminosity.

\subsection{Luminosity function}\label{S:lf}
\begin{figure*}
    \centering
    \includegraphics[width=2\columnwidth]{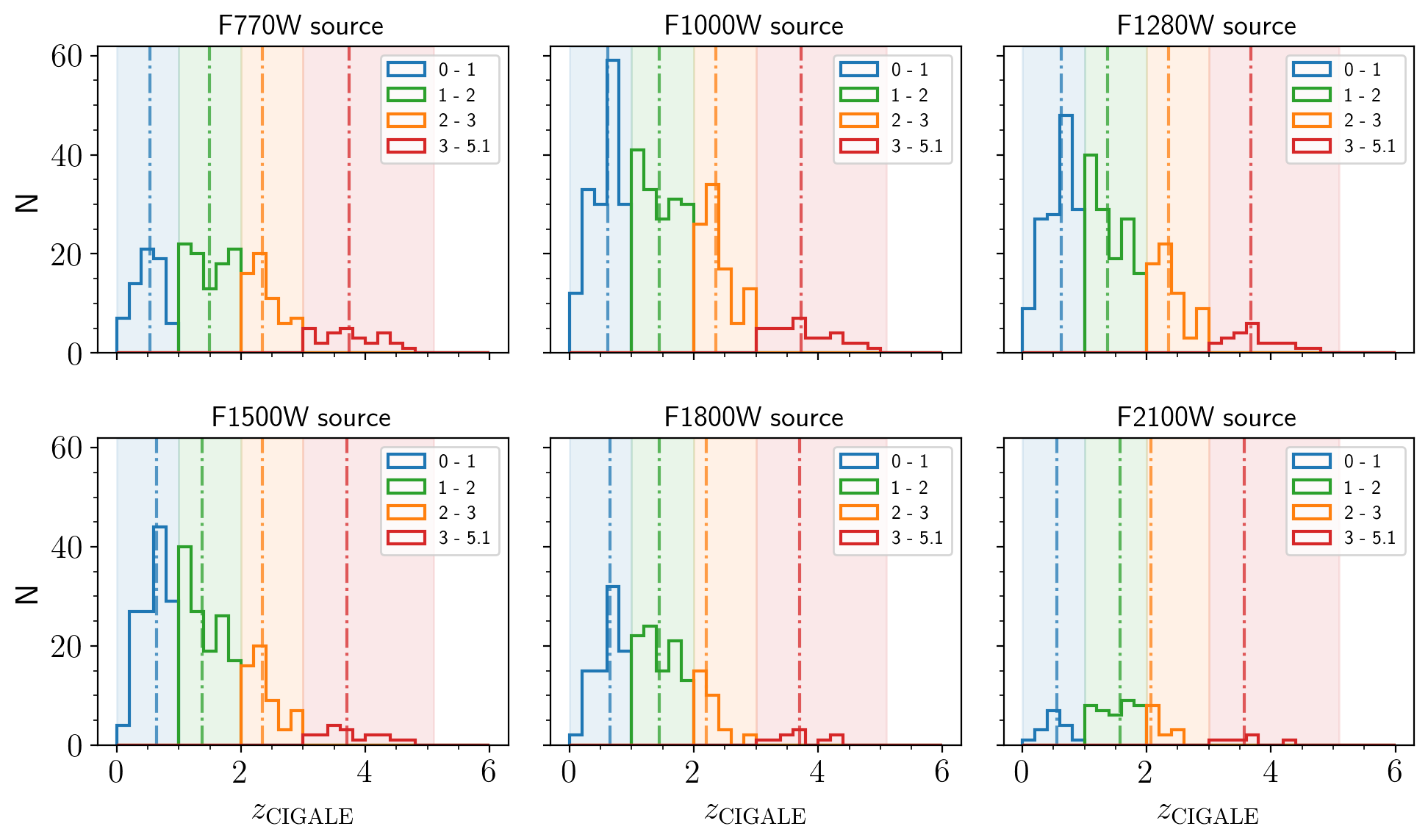}
    \caption{The redshift ($z_{\rm CIGALE}$) histogram of the sources in each band, divided into our redshift bins. The colour regions indicate the span of each bin, and the dot-dashed lines are the median.}
    \label{fig:z_hist}
\end{figure*}

The LF describes the number density of galaxies $\phi(L)$ as a function of their intrinsic brightness $L$ in a specific volume $V$. This volume depends on the redshift distribution of the galaxies, i.e.,
\begin{equation}\label{eq:v}
    V = V_C(z_{\max}) - V_C(z_{\min})
\end{equation}
where $V_C$ is the comoving volume, $z_{\min}$ and $z_{\max}$ are the minimum and maximum redshift for a specific redshift range.
It is conventional to refine the redshift range of the sample into multiple bins to investigate the evolution of the LF. In our case, four redshift bins, $z=[0,1], [1,2], [2,3], [3,5.1]$, are adopted as shown in Figure \ref{fig:z_hist}. Due to the insufficient sample size, a relatively large redshift bin of $z=[3,5.1]$ is chosen. We should be careful that the bin may include a wide range of variations in evolution. 

For a volume-limited sample, $V$ is just the survey volume. However, the survey is often limited by the sensitivity of the telescope. In practice, this results in a flux-limited sample, which can be incomplete in terms of volume.
An example is that the survey volume of redshift bin $z=[1,2]$ cannot be applied to faint galaxies with luminosities below a certain level, because they can only be detectable within $z<1.5$ for the filter.

The $1/V_{\max}$ method \citep{1968ApJ...151..393S} is thus introduced to address the volume incompleteness from a flux-limited sample. This method utilised the $z_{\max}$, the maximum redshift where a given object would still be detectable by the telescope, by redshifting the SED of the object toward the flux limit of a specific filter. 
The final $z_{\max}$ is taken as the redshift that reaches the flux limit. If the $z_{\max}$ goes out of the redshift bin, we use the upper limit of the redshift bin as $z_{\max}$.
Then, the effective survey volume $V_{\max}$ for the object can be calculated using equation \ref{eq:v}.
The exact formula for deriving the LF based on waveband $\nu$ is given by
\begin{equation}\label{eq:LF}
    \phi(L) = \frac{1}{\Delta \log L}\sum_i \frac{1}{V_{\max, i}} w_{i, \nu} \times \text{Compensation}
\end{equation}
where $\Delta \log L = 0.5$ dex is the width of our luminosity bin, and $w_{i, \nu}$ is the correction factor for $i$-th galaxy at a waveband $\nu$. Specifically, 
\begin{equation}
    w_{i, \nu} = \frac{4 \pi \, {\rm sr}}{{\rm area}_{\nu}} \times \frac{{\rm reliability}_{i, \nu}}{{\rm completeness}_{i, \nu}(F_{i, \nu})}
\end{equation}
area$_\nu$ is the sum of the sky coverage in every pointing of a waveband $\nu$ (see \S \ref{s:miri}). 
To obtain the completeness for a given galaxy $i$, we first select the completeness function corresponding to the specific waveband $\nu$ and pointing position of the galaxy, as shown in Figure \ref{fig:completeness}. 
Subsequently, the completeness can be computed from the $i$'s flux at waveband $\nu$ ($F_{i, \nu}$).
A similar approach is applied for the reliability correction, where the correction factor is also set according to the galaxy's pointing and waveband (Table \ref{tab:spurious}).
Notably, we derive TIR LF using corrections for F1000W band, because we find the largest sample of galaxies at 10 $\mu$m (see Figure \ref{fig:z_hist}).
In addition to the correction for individual galaxies, we add a global compensation factor of $\sim 1.13 \, (13\%)$ to the LF in each redshift bin for removed galaxies, as indicated in \S \ref{S:sed}.

Identifying limiting luminosities to filter out galaxies with insufficient completeness is crucial when interpreting LFs.
To achieve this, we have converted the 80\% flux completeness limits from Table \ref{tab:limit} into limiting luminosities. 
Two types of SED, SF and AGN galaxies, are assumed in the calculation based on the classification stated in \S \ref{S:sed}.
The SED of NGC6090 and Sey2 (representative of average Seyfert 2 galaxies) from SWIRE Template Library \citep{Polletta2007} are utilised as template SEDs for SF and AGN galaxies, respectively. These templates are selected according to \cite{Yang2023ApJ...950L...5Y} which suggests that the median SEDs of CEERS MIRI SF and AGN galaxies resemble them respectively.
The limiting luminosity is tailored to the redshift, completeness limit, and SED type for each galaxy by the assumed template SED.
In practice, we first obtain the 80\% completeness limit specific to the LF of the given waveband and the pointing of the galaxy, similar to the correction procedure above. 
Next, an appropriate template SED based on the galaxy's type (SF or AGN) is selected and redshifted to the desired redshift. 
We then integrate for the luminosity (either monochromatic or TIR, depending on the LF) at which the flux of SED at the specific band would be equal to the completeness limit as the limiting luminosity.
Galaxies with luminosities that fall below these limiting thresholds are regarded to be unreliable and are thus excluded.

\begin{figure}
    \centering
    \includegraphics[width=\columnwidth]{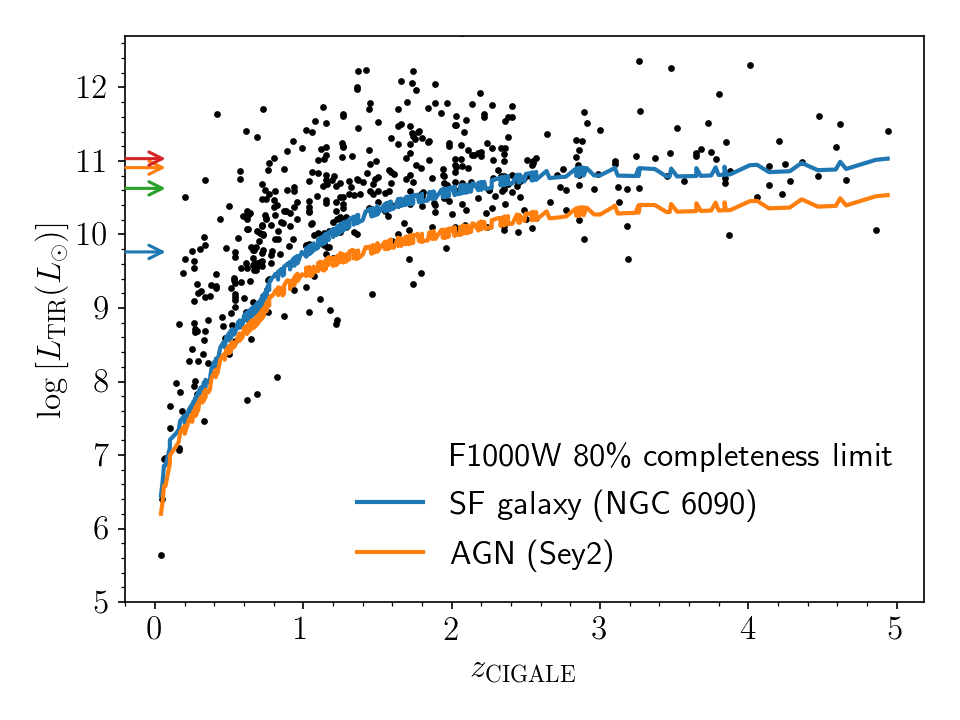}
    \caption{TIR luminosity ($L_{\rm TIR}$) as a function of redshift ($z_{\rm CIGALE}$). The solid line is the completeness luminosity limit converted from 80\% completeness flux limits for the F1000W band, assuming SF (blue) and AGN (orange) SED template. Galaxies below the line (depending on their SED type) are excluded from the analysis. The coloured arrows on the left indicate the TIR luminosity limits for each redshift bin used in this work, which is the brightest among all limits for different SED types in that redshift bin.}
    \label{fig:1000_lum_limit}
\end{figure}

Figure \ref{fig:1000_lum_limit} shows the criteria for the TIR LFs (Figure \ref{fig:TIR_LF_med}) which uses the F1000W band limit.
We note the small zigzags in these completeness lines simply reflect the slight difference in flux completeness limits for the galaxies at each CEERS pointing. 

The final size of our galaxy samples for monochromatic and TIR LFs after the 80\% completeness selection is listed in Table \ref{tab:count}.
We derived luminosity limits for all redshift bins $z=[0,1], [1,2], [2,3], [3,5.1]$ in each LF. These limits are determined by the brightest among all limiting luminosities of different SED types (i.e., SF and AGN) assumed in the given redshift bins, as illustrated in Figure \ref{fig:1000_lum_limit}. The luminosity limits are shown along with the corresponding LFs in \S\ref{S:mono_LF} and \S\ref{s:TIR} to help readers confirm at what luminosity the LFs are confident.

\begin{table}
    \centering
    \begin{threeparttable}[b]
    \begin{tabular}{cccc}
    \hline
     & \multicolumn{3}{c}{\textbf{Population}} \\
                       & All\tnote{a}         & SF\tnote{b}          & AGN\tnote{c}         \\ \hline
    F770W limit        & 181         & 126         & 55         \\
    F1000W limit       & 444 (409)\tnote{d}      & 320         & 124         \\
    F1280W limit       & 316         & 226         & 90         \\
    F1500W limit       & 300         & 215         & 85          \\
    F1800W limit       & 180         & 137         & 43          \\
    F2100W limit       & 45         & 35         & 10          \\ \hline
    \end{tabular}
    \begin{tablenotes}
        \item \textbf{Refer to:}
        \item[a] Figure \ref{fig:mono_6LFs} (monochromatic LFs for all galaxies)
        \item[b] Figure \ref{fig:mono_6LFs_SF} (monochromatic LFs for SF galaxies)
        \item[c] Figure \ref{fig:mono_6LFs_AGN} (monochromatic LFs for AGN)
        \item[d] Figure \ref{fig:TIR_LF_med} (TIR LFs for all galaxies)
    \end{tablenotes}
    \end{threeparttable}
    \caption{The final sample size for monochromatic and TIR LFs based on 80\% completeness limit selection in each filter and galaxy population.}
    \label{tab:count}
\end{table}

\section{Results and discussion}\label{S:res}
\subsection{Monochromatic LFs}\label{S:mono_LF}
We present the monochromatic rest-frame LFs at 7.7, 10, 12.8, 15, 18 and 21 $\mu$m, based on 6 available MIRI filters from \textit{JWST}. 
In \S \ref{s:all_mono_lf} we show the LFs of all galaxy populations, and in \S \ref{S:SF_AGN} we discuss the LFs for SF and AGN host galaxies.
To estimate the error, we resample the LF distribution 100 times. This is done by perturbing the redshift (i.e. photo-$z$) of each galaxy with associated probability distributions from \textsc{cigale} and then recalculating the luminosities (monochromatic and TIR), as well as corresponding LFs. Subsequently, we add a Poisson error to each luminosity bin.
Bins with less than 3 galaxies are removed to avoid fluctuations from small statistics.

For comparison, we overplot the LFs from previous observations with IR space telescopes. We provide a brief summary of them below.
\cite{2019PASJ...71...30G} and \cite{2015MNRAS.454.1573K} are based on \textit{AKARI} observations in the 5.4 deg$^2$-wide NEP field. \cite{2019PASJ...71...30G} measures the rest-frame 8 and 12 $\mu$m LFs at $0.35<z<2.2$ using 18 bands mid-IR photometry \citep{Kim2012A&A...548A..29K} and 5 optical bands from the Subaru Hyper Suprime-Cam \citep{Goto2017PKAS...32..225G}, while \cite{2015MNRAS.454.1573K} focuses on local ($z<0.3$) LFs at 8, 12 and 15 $\mu$m with a spectroscopic only sample.
\cite{2005ApJ...632..169L}, \cite{2005ApJ...630...82P}, \cite{2006MNRAS.370.1159B} and \cite{2010A&A...515A...8R} utilise observations from \textit{Spitzer}. 
\cite{2005ApJ...632..169L} derives rest-frame 15 $\mu$m LF at $z \lesssim 1$ and \cite{2005ApJ...630...82P} derives rest-frame 12 $\mu$m LF to $z\simeq 2.6$ based on the 24 $\mu$m selected sample from the CFDS field.
\cite{2006MNRAS.370.1159B} builds rest-frame 8 and 24 $\mu$m LFs at $0<z<2$ with one SWIRE field. With the deep observations in the VVDS-SWIRE field, \cite{2010A&A...515A...8R} presents rest-frame 8, 12, 15, and 24 $\mu$m LFs up to $z\simeq 2.5$. We only plot LFs from the literature whose redshift ranges are similar to the median of our redshift bins.

We first remark on the superior sensitivity of the \textit{JWST}. In Figure \ref{fig:mono_6LFs}, we find the luminosities of galaxies can be probed down to $L^* \sim 10^7 - 10^8 L_\odot$ for the lowest redshift bin $z=0-1$, while their luminosity limits are at $L^* \sim 10^8 - 10^9 L_\odot$ at higher redshifts. Still, This is about 1 to 2 orders of magnitude fainter than the limiting luminosities found by \textit{AKARI} and \textit{Spitzer} at similar redshifts ($\sim 10^9 - 10^{10} L_\odot$). 
It is no surprise as such improvement has been shown in early studies on MIR data from \textit{JWST} \citep[e.g.,][]{2022MNRAS.517..853L, Wu2023MNRAS.523.5187W}. 

\subsubsection{LFs of all galaxy populations}\label{s:all_mono_lf}
\begin{figure*}
    \centering
    \includegraphics[width=1.85\columnwidth]{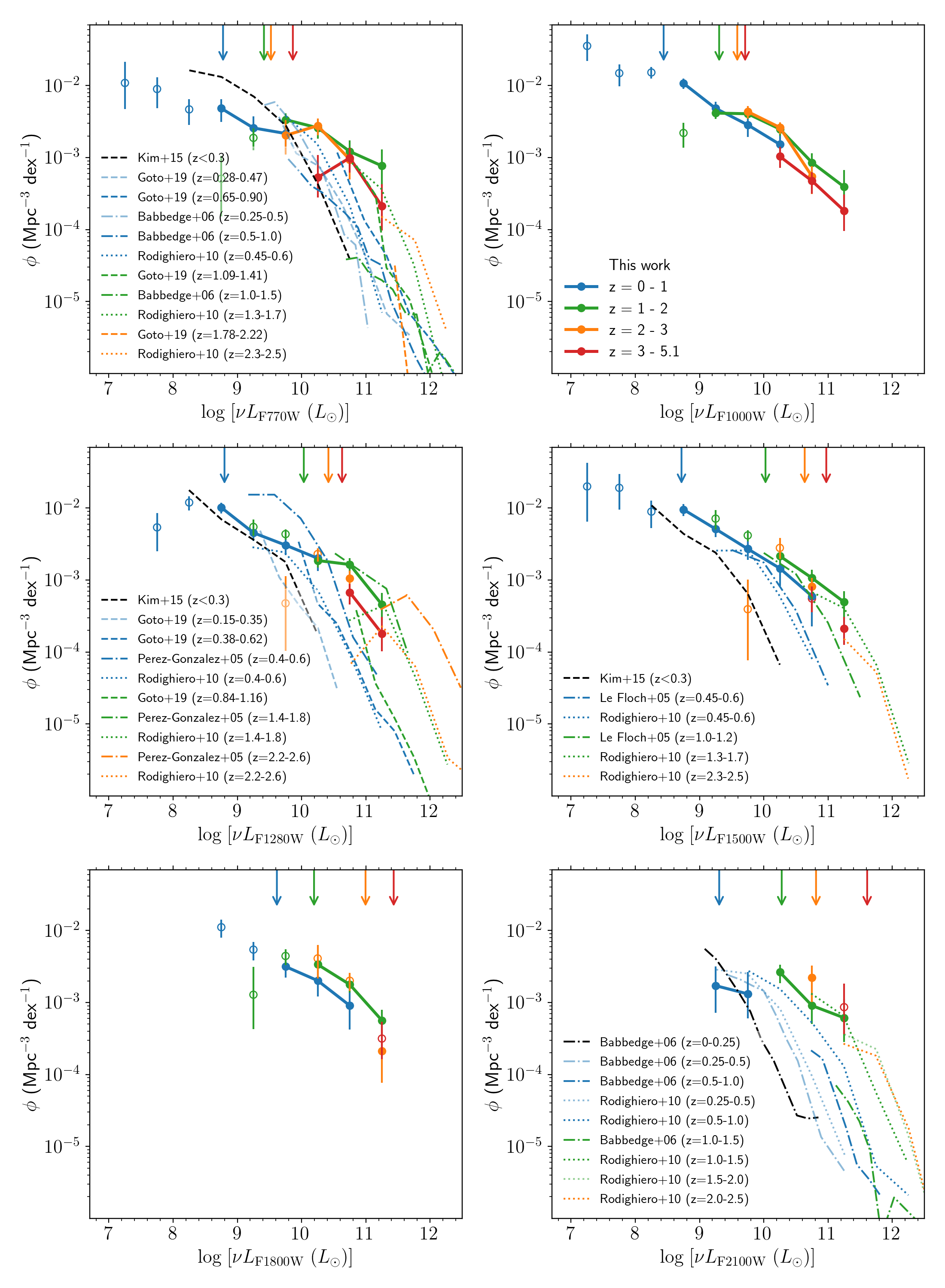}
    \caption{The 7.7 (F770W), 10 (F1000W), 12.8 (F1280W), 15 (F1500W), 18 (F1800W) and 21 (F2100W) $\mu$m monochromatic rest-frame luminosity ($\nu L_\nu$) functions. The four redshift bins $z=0-1$ (blue), $z=1-2$ (green), $z=2-3$ (orange), and $z=3-5.1$ (red) are shown by circles. Luminosity limits for each redshift bin are shown by coloured arrows at the top.
    The data points below the limits are shown as open markers and are not connected.
    For 7.7, 12.8, 15, and 21 $\mu$m, LFs from the literature \protect\citep{2019PASJ...71...30G, 2005ApJ...632..169L, 2005ApJ...630...82P, 2006MNRAS.370.1159B, 2010A&A...515A...8R} using redshift bins close to ours are also plotted. These LFs are represented with different line styles (dashed/dot-dashed/dotted), while the same colour indicates the same redshift bin. The black dashed/dot-dashed line shows the local LF from \protect\cite{2015MNRAS.454.1573K} and \protect\cite{2006MNRAS.370.1159B}.}
    \label{fig:mono_6LFs}
\end{figure*}

In the four panels (7.7, 12.8, 15 and 21 $\mu$m) in Figure \ref{fig:mono_6LFs} that have previous results overplotted, our LFs agree with and well extend the faint end from previous works to more than one order of magnitude. 
The luminosity bins in our study show little overlap with those reported in the literature, which implies that most parts of our LFs are observed for the first time, especially for $z>2$. 
Among all the monochromatic LFs, a gradual evolution in luminosity can be seen clearly among all redshift bins. At higher redshift ($z>3$), the change in LF curves is dominated by the density evolution.
The overall evolution between $z=1-2$ and $z=2-3$ appears more moderate.
At high-$z$ ($>1$) our data are incomplete at low luminosities, as indicated by coloured arrows. Therefore, the faint end of the LF is not sampled and cannot be estimated. 

\subsubsection{LFs of SF and AGN galaxies}\label{S:SF_AGN}
\begin{figure*}
    \centering
    \includegraphics[width=1.85\columnwidth]{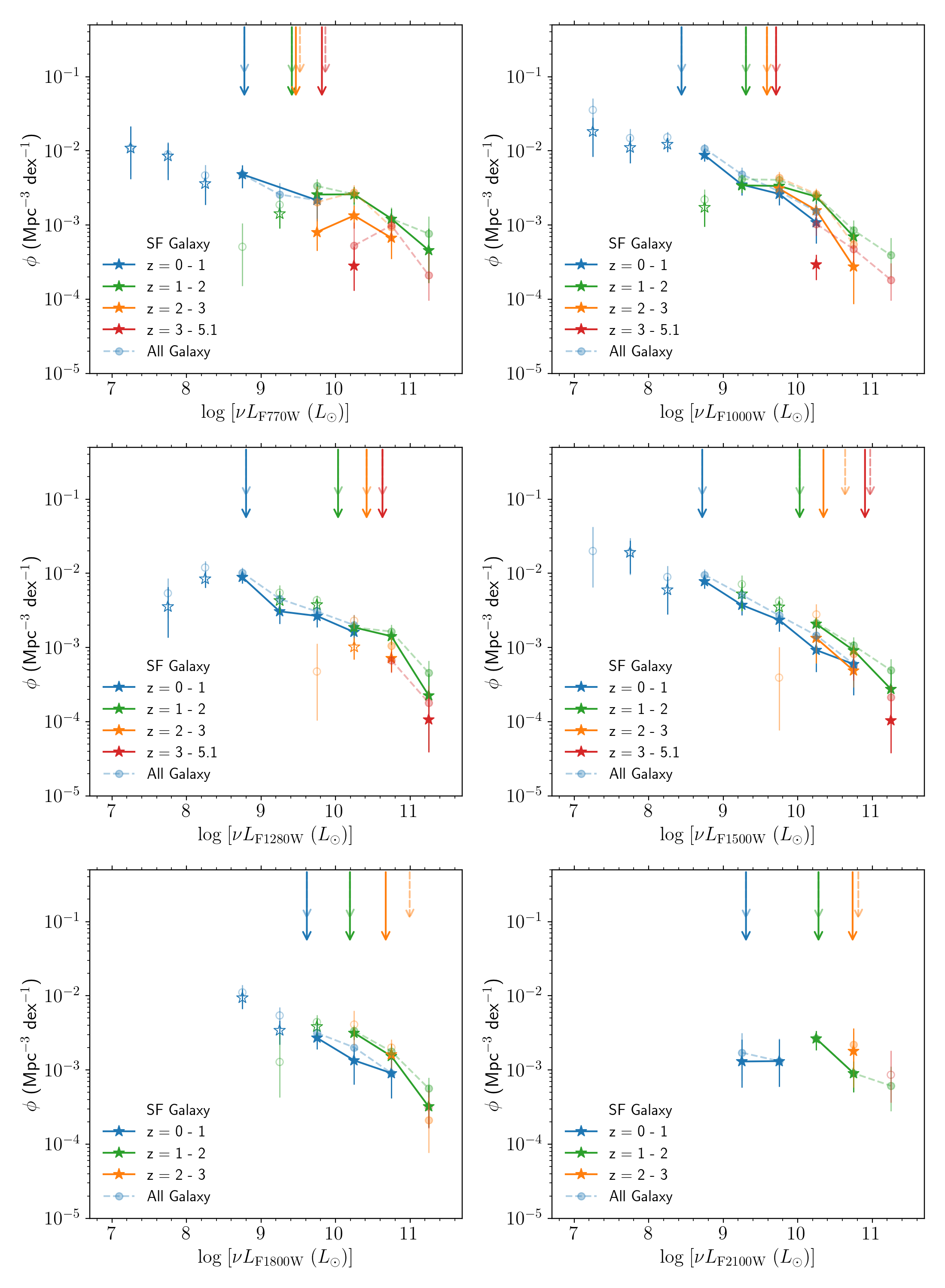}
    \caption{Same as Figure \ref{fig:mono_6LFs} but for star-forming (SF) galaxies (shown in star / solid line) only. Galaxy LFs for all populations from Figure \ref{fig:mono_6LFs} are also plotted (shown in circle / dashed line) for reference. We also provide luminosity limits for SF galaxies (longer solid arrows) and all galaxies (shorter dashed arrows, obtained from Figure \ref{fig:mono_6LFs}).}
    \label{fig:mono_6LFs_SF}
\end{figure*}

\begin{figure*}
    \centering
    \includegraphics[width=1.85\columnwidth]{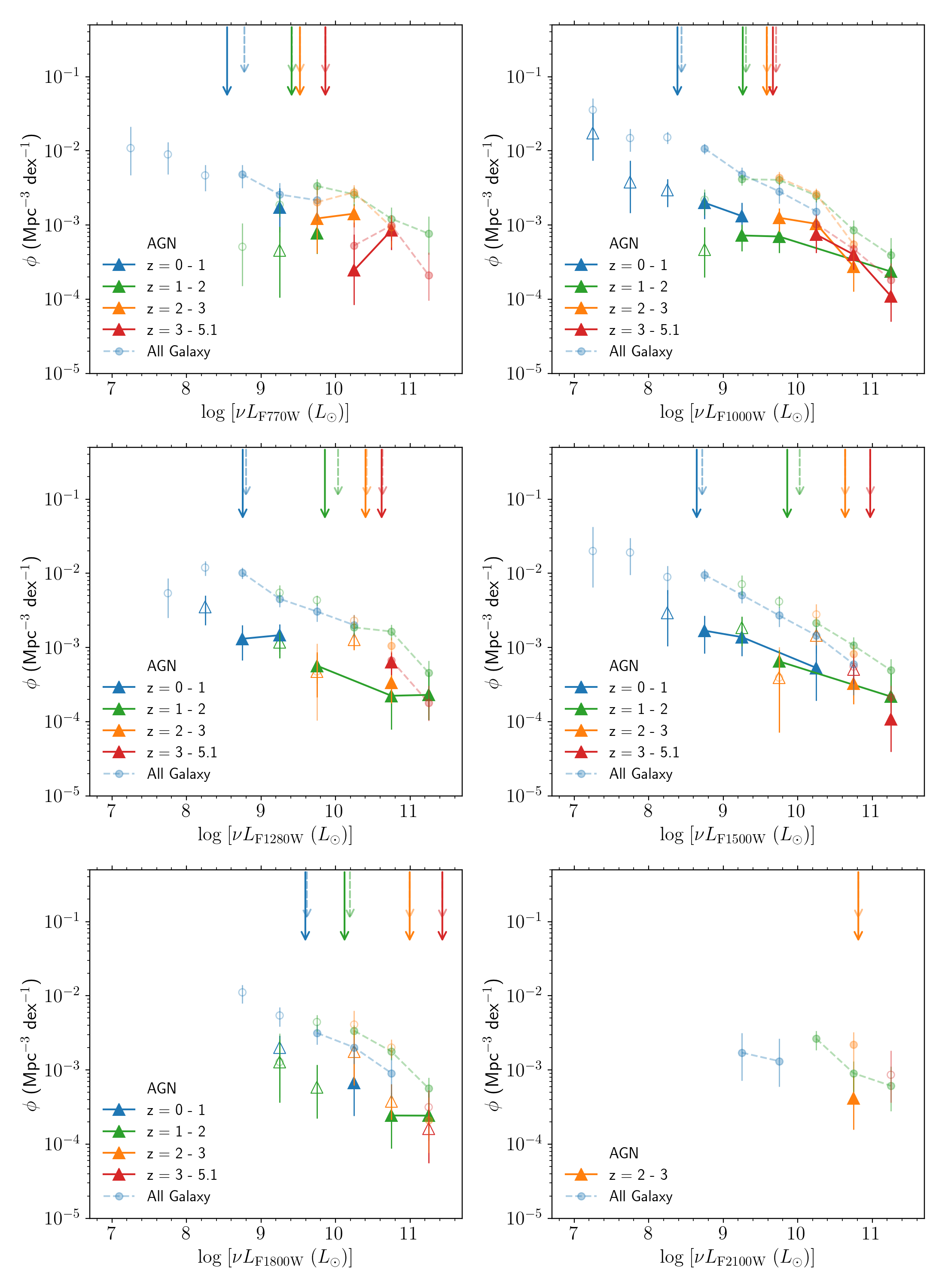}
    \caption{Same as Figure \ref{fig:mono_6LFs_SF} but for AGN host galaxies (shown in triangle / solid line) only. We also provide luminosity limits for AGNs (longer solid arrows) and all galaxies (shorter dashed arrows, obtained from Figure \ref{fig:mono_6LFs}).}
    \label{fig:mono_6LFs_AGN}
\end{figure*}

While MIR emission serves as a good proxy for star-formation activity, it may be contaminated by the radiation from heated dust due to AGN activity. 
To study the impact of AGN on LF from different galaxy populations, we obtain LF for typical SF and AGN host galaxies separately by frac$_{\rm AGN}$ (refer to \S \ref{S:sed}). The resulting LFs are presented in Figure \ref{fig:mono_6LFs_SF} (SF) and \ref{fig:mono_6LFs_AGN} (AGN). LF for all galaxies (from Figure \ref{fig:mono_6LFs}) are also overplotted here for reference.

We find that LFs from SF galaxies generally follow those from all galaxies, exhibiting only slight differences ($\lesssim 0.2$ dex), as the majority of galaxies in our sample are SF galaxies. The main features of LFs in Figure \ref{fig:mono_6LFs} (all galaxies) can be also found in Figure \ref{fig:mono_6LFs_SF} (LFs of SF galaxies).
The disappearance of the high-$z$ bright end in Figure \ref{fig:mono_6LFs_SF} (due to the removal of luminous AGN) implies a potentially stronger evolution in SF galaxies compared to what is suggested in Figure \ref{fig:mono_6LFs}, because fewer SF galaxies are expected to be found at $z>3$.

Figure \ref{fig:mono_6LFs_AGN} suggests that AGN host galaxies are less significant at lower redshift within our main luminosity coverage $L^* \sim 10^8-10^{11} L_\odot$. AGNs are more dominant at the highest redshift bin at $z=3-5.1$, as presented by previous analyses on IR populations \citep[e.g.,][]{2013MNRAS.432...23G}. 
Nevertheless, it is difficult to draw a solid conclusion on the AGN evolution within the error bar, because of the insufficient sample size.

\subsection{TIR LF}\label{s:TIR}
\begin{figure}
    \centering
    \includegraphics[width=\columnwidth]{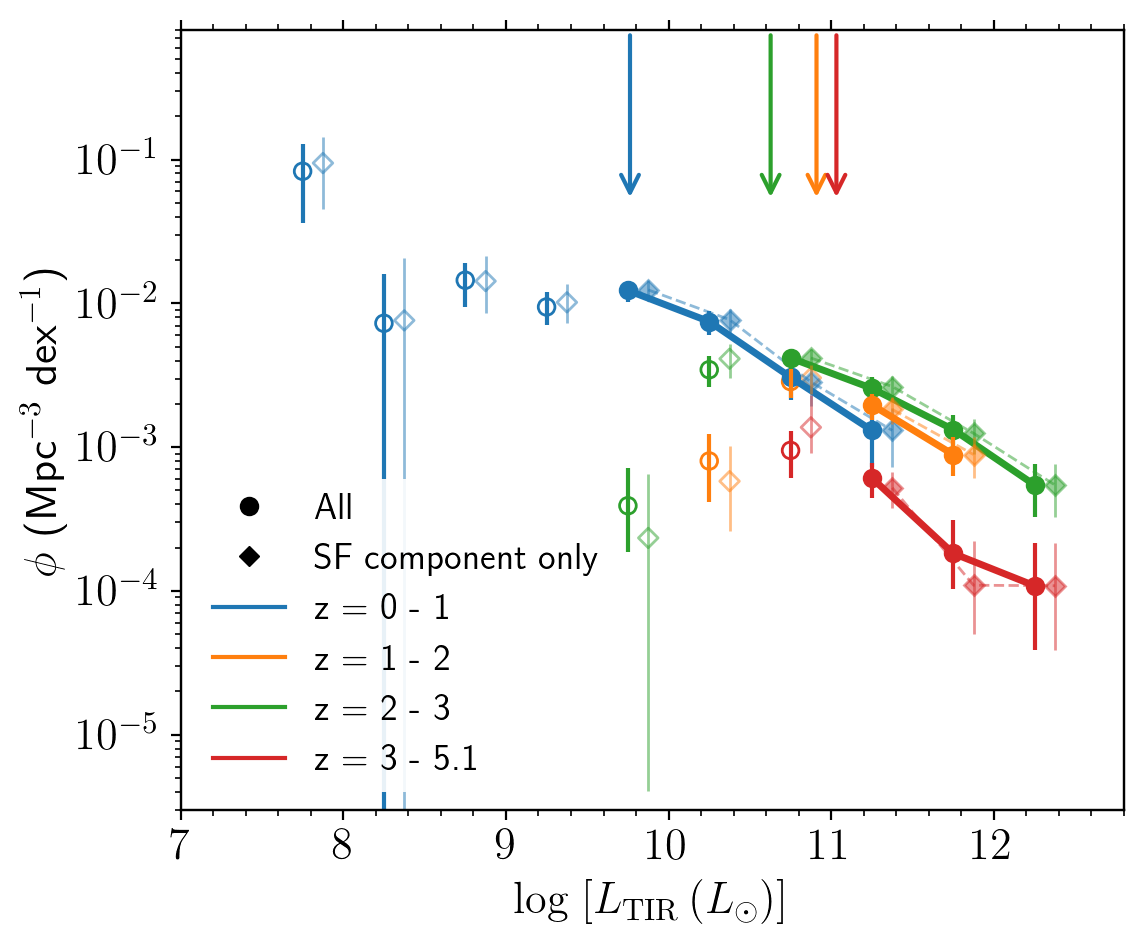}
    \caption{The rest-frame total infrared (TIR) luminosity functions. The four redshift bins $z=0-1$ (blue), $z=1-2$ (green), $z=2-3$ (orange), and $z=3-5.1$ (red) are shown. 
    TIR LFs derived from all galaxies (including AGN components in the SEDs) are marked by circles. For comparison, TIR LFs derived from using only star-forming components of the SED (i.e., excluding the contribution from AGN components of the SEDs such as torus, polar dust, and disk emissions) are marked by diamonds, with an offset of 0.125 dex in $L_{\rm TIR}$ for clarity. 
    The coloured arrows at the top show the luminosity completeness limits, adopted from Figure \ref{fig:1000_lum_limit}.
    The data points below the limits are shown as open markers and are not connected.}
    \label{fig:TIR_LF_med}
\end{figure}

We now show the total (bolometric) IR LF in Figure \ref{fig:TIR_LF_med}. The procedure to construct the TIR LF is the same as what has been described in \S \ref{S:mono_LF}, and we use the F1000W limit-selected sample (refer to Figure \ref{fig:1000_lum_limit} and Table \ref{tab:count}). 409 galaxies in total are used to obtain the TIR LF regardless of galaxy type, i.e., all galaxies are used including AGN components, indicated by circles. 
For comparison, we also show the TIR LFs which only use star-forming components of the SEDs. AGN components of the SEDs (torus, polar dust, and disk emissions) are excluded in computing those LFs shown with diamond symbols, in order to examine the possible bias imposed by AGNs in interpreting star-forming activities.
No significant difference is found in comparing TIR LFs that include all the SED components (circles) with those excluding AGN components (diamonds). Their medians are only $<6\%$ different on average. This suggests that AGN emissions do not dominate the SED in the far-IR wavelengths. Note that it differs from our frac$_{\rm AGN}$ definition, which only depends on emission within $3-30$ $\mu$m. 
As a result, the impact from the AGN components is expected to be marginal.

Comparing Figure \ref{fig:TIR_LF_med} to Figure \ref{fig:mono_6LFs_SF}, we find the evolution of TIR LF traces the trend of 7.7 and 12.8 $\mu$m SF galaxy LFs. The correlation between 7.7 and 12.8 $\mu$m luminosity and TIR luminosity results from the strong PAH emission at these wavelengths, which plays a significant role in shaping the IR spectrum of SF galaxies. We note that this has been investigated in previous studies \citep[e.g.,][]{Caputi2007, 2008A&A...479...83B, 2011MNRAS.410..573G, Lin2023}.

While we have pushed the TIR LF to a notably high-$z \sim 5$ with the brand new \textit{JWST} data, we should caution that the TIR luminosity of galaxies at higher redshift ($z>3$) could be undetermined. The dust emission in MIR which we trace for deriving TIR luminosity will start to be replaced by stellar emission at high-$z$, thus the FIR (and partially MIR) part of SED will mostly rely on extrapolation of the model. 
However, with the reddest MIRI bands in the CEERS field (F1800W and F2100W), 3.3 $\mu$m PAH features from star-forming activity can still be detected at $z=4-5$, as discussed in \cite{Yang2023ApJ...950L...5Y}. Furthermore, the majority of such high-$z$ galaxies are AGN (refer to \S \ref{S:SF_AGN}). When considering MIRI coverage, the SEDs of these galaxies (Figure \ref{fig:sed_example_02} and \ref{fig:sed_example_03}) are typically dominated by AGN emission rather than stellar emission, which indicates a clue to the FIR SED. Previous studies \citep[e.g.,][]{Elbaz2011A&A...533A.119E, Dai2018MNRAS.478.4238D} have shown a redshift-independent relation between AGN emission and FIR / TIR luminosity. 
Besides these points, we caution readers about possible uncertainty in estimating TIR luminosity based on the MIRI flux.

\subsubsection{Comparison with the literature}
Figure \ref{fig:TIR_LF_zbin_00}-\ref{fig:TIR_LF_zbin_03} compare our LFs to the literature, separated in each redshift bin. The luminosity completeness limits for our work and literature are denoted by different markers. Fitted curves of our LF are also presented, refer to Section \ref{S:MCMC}.
Both \cite{2013MNRAS.432...23G} (open/closed coloured circles) and \cite{2013A&A...553A.132M} (coloured stars) use the FIR observations from \textit{Herschel} to produce TIR LF. Specifically, \cite{2013MNRAS.432...23G} investigated an extensive field containing the GOODS, ECDFS and COSMOS area, and \cite{2013A&A...553A.132M} concentrated on the deep pencil beam GOODS-S field. Both of them have similar luminosity limits with respect to their redshift range, as the markers suggest. Note that we derive the limits for \cite{2013MNRAS.432...23G} with the nominal 100 $\mu$m limiting flux of SF-AGN galaxy in the GOODS-S field (1.2 mJy), assuming the SED of NGC6090. 
\cite{Gruppioni2020A&A...643A...8G} (open diamonds) utilise the sub-mm observations from the ALMA ALPINE survey. By tracing the rest-frame FIR continuum emission, they first extend the IR LF to a wide range $z=0.5-6$.

Similar to the monochromatic LFs, our results are consistent with previous studies and further advance them more than one order of magnitude fainter. The depth of \textit{JWST} enables us to explore the faintest MIR objects ever seen, with a luminosity limit roughly corresponding to $L^* \sim 10^{9.7} L_\odot$ ($z=0-1$), $10^{10.6} L_\odot$ ($z=1-2$), $10^{10.9} L_\odot$ ($z=2-3$), and $10^{11} L_\odot$ ($z=3-5.1$).
To put this into context, these limits are approximately 1.5 dex fainter compared to the previous MIR space telescope \textit{AKARI} \citep{2019PASJ...71...30G}. In all the redshift bins, we extend the limits of \textit{Herschel} works by an order of magnitude. Additionally, for higher redshifts, we push the luminosity bins to $\sim 0.5$ dex fainter than those derived from \cite{Gruppioni2020A&A...643A...8G}.

The shape of \textit{JWST} TIR LF is similar to the \textit{Herschel} or \textit{AKARI} works. Still, we would like to point out a deviation compared to \cite{Gruppioni2020A&A...643A...8G} at $z=1-2$ and $z=2-3$, where our LFs are higher and steeper in the faint end. 
The deviation has been previously suggested in the comparison with \cite{2013MNRAS.432...23G} and \cite{2013A&A...553A.132M} in \cite{Gruppioni2020A&A...643A...8G}, though it was not significant according to their luminosity limits. 
Given that we now constrain the LF to a much fainter luminosity (for their $z=1.5-2.5$ and $z=2.5-3.5$ bins), this deviation is not negligible, although we must note that the redshift bins are not exactly the same (i.e., those of \citealt{Gruppioni2020A&A...643A...8G} are shifted up by about $dz=0.5$). On the other hand, we observe a good agreement between our data points and \cite{Gruppioni2020A&A...643A...8G} for the highest redshift. 

\subsubsection{MCMC analysis}\label{S:MCMC}
To quantify the evolution of the TIR LFs, we adopt a modified-Schechter function \citep{Saunders1990} to fit the LFs:
\begin{align}\label{eq:fit}
    \phi(L)d \log L &= \phi^* \left( \frac{L}{L^*} \right)^{1-\alpha} \exp \left[ -\frac{1}{2\sigma^2} \log^2_{10} \left( 1+\frac{L}{L^*} \right) \right] d \log L
\end{align}
which is a power law (determined by $\alpha$) for $L<L^*$ and a Gaussian in $\log L$ (determined by $\sigma$) for $L>L^*$. $\phi^*$ is the normalised factor for density. 
We apply the Markov chain Monte Carlo (MCMC) method to fit our LFs. This is conducted by the Python package \textsc{emcee} \citep{EMCEE2013PASP..125..306F}, with 100 walkers initialised and 1000 iteration steps. Data points are allowed to move within their error bars in each iteration, assuming a Gaussian distribution. 
Given that our data points mostly probe the faint end of the whole LF, it is not sufficient to fit the four free parameters $L^*$, $\phi^*$, $\alpha$ and $\sigma$ simultaneously. To overcome the issue, we use a technique similar to \cite{2006MNRAS.370.1159B} and \cite{Gruppioni2020A&A...643A...8G} for fitting.

First, we fix the slope of the bright end $\sigma$ to 0.5. The value is obtained from the \textit{Herschel} LF \citep{2013MNRAS.432...23G} that has been well-constrained on bright galaxies $L^* = 10^{11}-10^{13} L_\odot$. Then, we fit the lowest redshift bin $z=0-1$ to the remaining three parameters. 
A flat prior range of $\log (L^*/L_\odot)=[8, 13]$, $\log (\phi^*/{\rm Mpc^{^3}dex^{-1}})=[-5, -1]$, and $\alpha=[-1,3]$ are set. 
We determined and fixed $\alpha$ from the fitting result, which means only $L^*$ and $\phi^*$ are fit for the rest of the redshift bins. 

\begin{figure*}
    \centering
    \includegraphics[width=1.15\columnwidth]{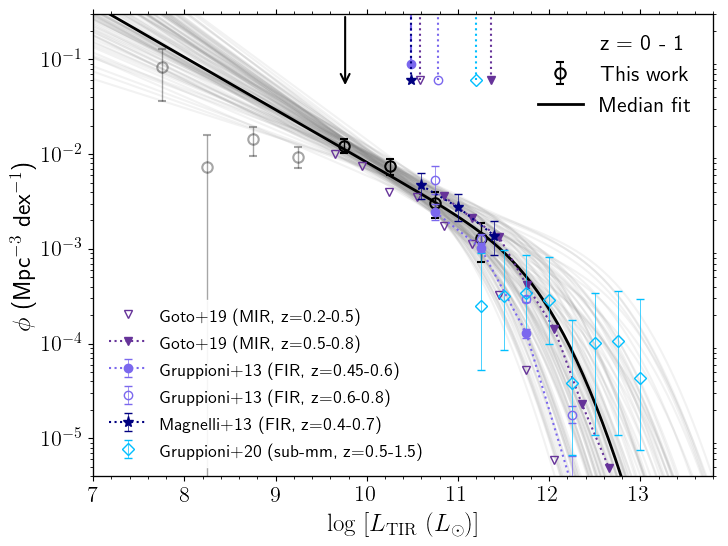}
    \includegraphics[width=.85\columnwidth]{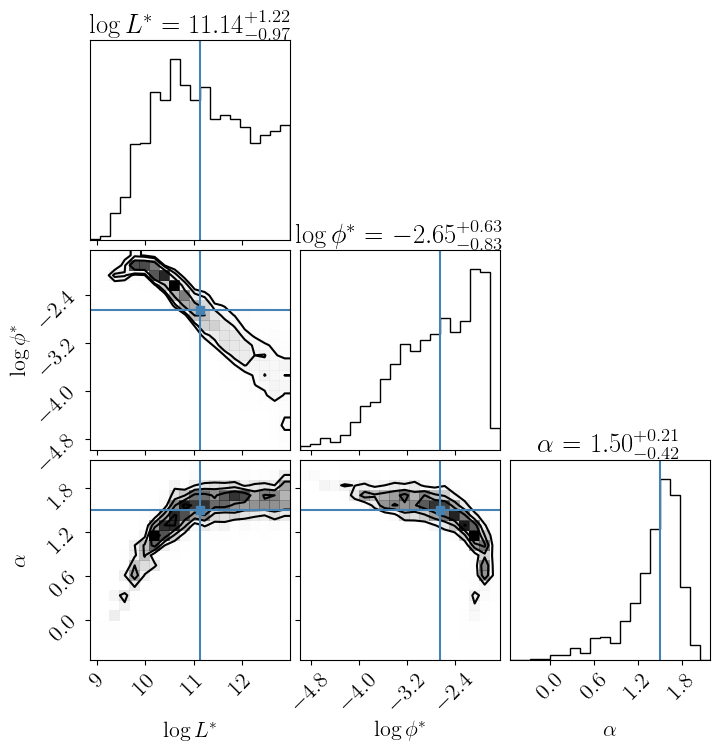}
    \caption{Left panel: the rest-frame total infrared (TIR) luminosity functions, same as Figure \ref{fig:TIR_LF_med} but for $z=0-1$ bin only. The TIR LF are marked in black open circles. Grey open circles show luminosity bins not used in fitting. The median model fits are plotted in black solid lines, and the grey lines are the fits within the 1-sigma uncertainty range of the parameters.
    TIR LFs from previous works \protect\citep{2013A&A...553A.132M, 2013MNRAS.432...23G, 2019PASJ...71...30G, Gruppioni2020A&A...643A...8G} with similar redshift range are overplotted.    
    Markers on the top indicate the luminosity completeness limits of literature with the same marker, where the black arrow is for our work. Right panel: the corner plot showing the probability distribution of fit parameters from MCMC analysis. The median (blue cross), as well as the 16th- and 84th percentile of fit parameters, are provided. }
    \label{fig:TIR_LF_zbin_00}
\end{figure*}

\begin{figure*}
    \centering
    \includegraphics[width=1.15\columnwidth]{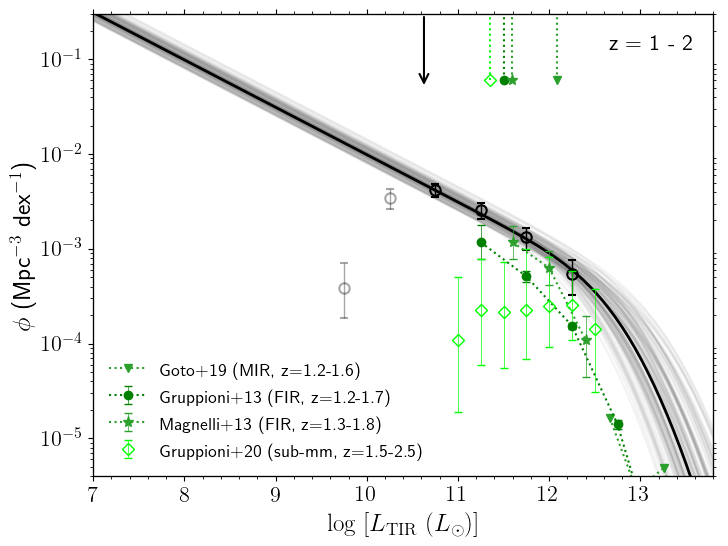}
    \includegraphics[width=.85\columnwidth]{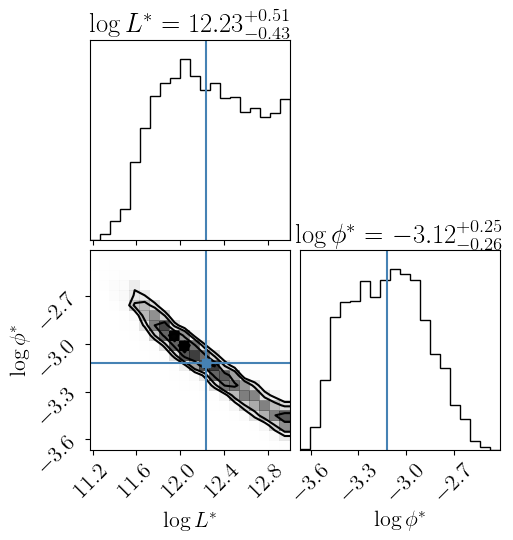}
    \caption{Same as Figure \ref{fig:TIR_LF_zbin_00} but for $z=1-2$ bin only.}
    \label{fig:TIR_LF_zbin_01}
\end{figure*}

\begin{figure*}
    \includegraphics[width=1.15\columnwidth]{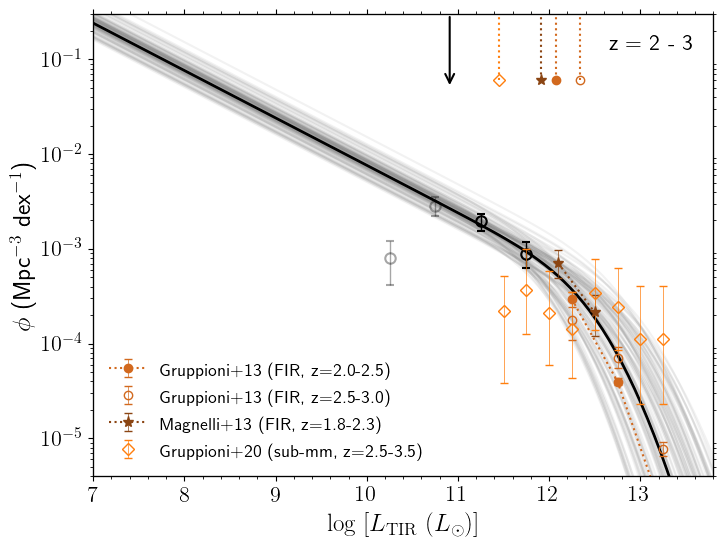}
    \includegraphics[width=.85\columnwidth]{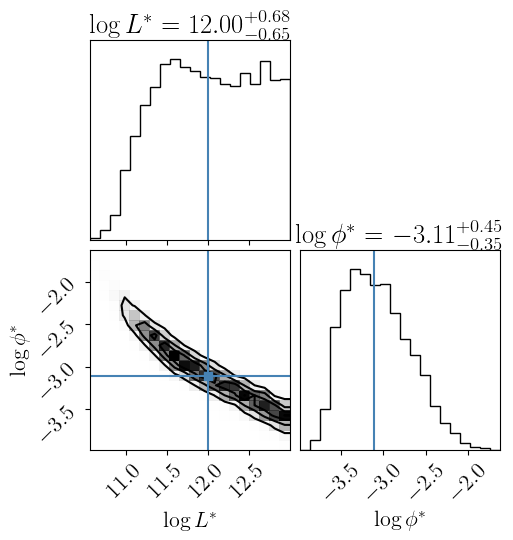}
    \caption{Same as Figure \ref{fig:TIR_LF_zbin_00} but for $z=2-3$ bin only.}
    \label{fig:TIR_LF_zbin_02}
\end{figure*}

\begin{figure*}
    \includegraphics[width=1.15\columnwidth]{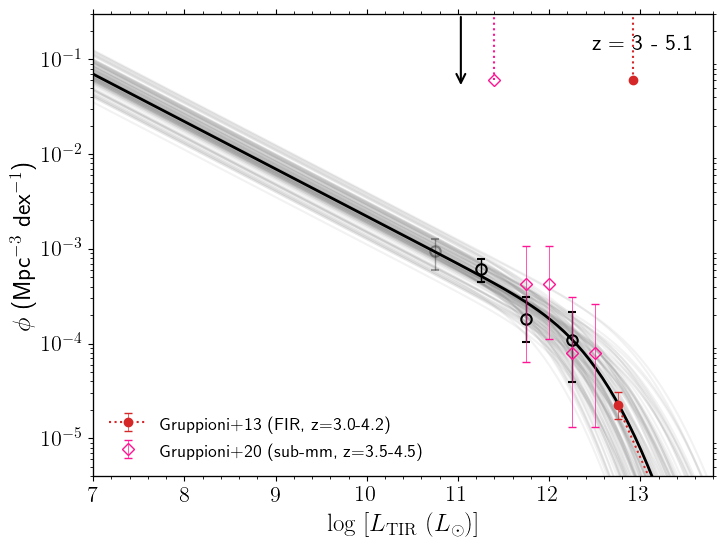}
    \includegraphics[width=.85\columnwidth]{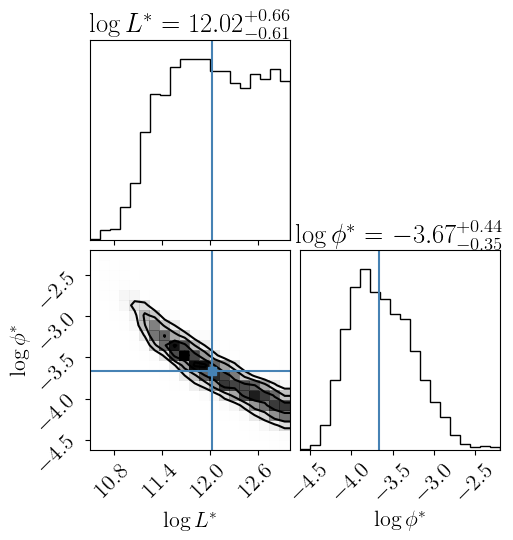}
    \caption{Same as Figure \ref{fig:TIR_LF_zbin_00} but for $z=3-5.1$ bin only.}
    \label{fig:TIR_LF_zbin_03}
\end{figure*}

The median and 1-sigma uncertainty for fitted parameters are summarised in Table \ref{tab:2}, where we use the 16th- and 84th percentile from the MCMC results for the uncertainty range. We overplot the fit curves within 1-sigma uncertainty (grey lines) and the median fit curve (black lines) in Figure \ref{fig:TIR_LF_zbin_00}-\ref{fig:TIR_LF_zbin_03} for each redshift bin. The median fit curve for pure SF TIR LF (dashed line) is also provided.
Faint luminosity bins that are further away from the luminosity completeness limit are excluded to avoid bad fits. These bins are shown in grey. In the right panel of Figure \ref{fig:TIR_LF_zbin_00}-\ref{fig:TIR_LF_zbin_03}, we show the probability distributions of the fit parameters. 

\begin{figure}
    \centering
    \includegraphics[width=\columnwidth]{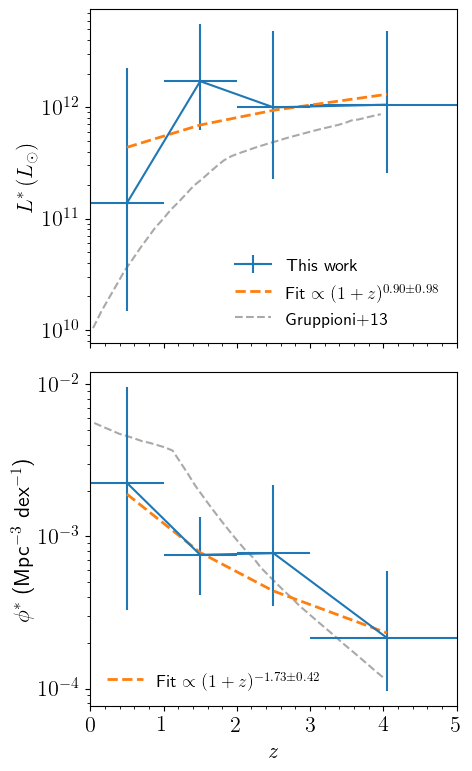}
    \caption{The redshift evolution of $L^*$ and $\phi^*$ from the fitted TIR LF, taken from Table \ref{tab:2}. The horizontal error bar is the redshift range of the bin. The model fit from \protect\cite{2013MNRAS.432...23G} is provided (grey dashed line), and our best fits for the data points are shown in orange dashed lines.}
    \label{fig:TIR_LF_evolution}
\end{figure}

The MCMC analysis suggests a strong degeneracy in $L^*$ and $\phi^*$. While the slope of the faint end $\alpha$ is relatively constrained (right panel of Figure \ref{fig:TIR_LF_zbin_00}), parameters for the knee of the LF seem to be highly undetermined, especially for $L^*$, due to the limited samples with high luminosity. We further address this issue in \S \ref{S:evolute}.
From the weakly constrained $\phi^*$, we also notice a clear scaling relation between $L*$ and $\phi^*$. 
The density ($\phi^*$) and luminosity ($L^*$) evolution from our median-fit TIR LF are shown in Figure \ref{fig:TIR_LF_evolution}. 
In addition, we overplot the best-fit curves, which parameterised the evolutions as $L^* \propto (1+z)^{0.90\pm0.98}$ and $\phi^* \propto (1+z)^{-1.73\pm0.42}$.
These curves are obtained by fitting the data points with non-linear least squares, assuming a function form of $a(1+z)^b$, where $a$ is the normalisation factor and $b$ is the slope. The associated errors are obtained from the derived covariance matrix. 
Our findings indicate that our fit broadly follows the fitted evolution presented in \cite{2013MNRAS.432...23G} but with a noticeably large error bar.
While $\phi^*$ indeed exhibits a decreasing trend with redshift, the overall evolution of $L^*$ appears to be minor, as it falls within the uncertainty range caused by significant degeneracy. 
Despite this, it is important to note that the global shape of fitted LF does not show obvious changes within the 1-sigma range (grey lines) due to the strong association between $L^*$ and $\phi^*$. Hence, we stress that such degeneracy is unlikely to impact the subsequent integration of luminosity density. 

\begin{table}
    \centering
    \begin{tabular}{|c|c|c|c|c|}
        \hline
        $z$ & $\log\,[L^*/L_\odot]$ & $\log\,[\phi^*/{\rm Mpc^{-3}dex^{-1}}]$ & $\alpha$ & $\sigma$\\
        \hline
        $0.50$ & $11.14\substack{+1.22 \\ -0.97}$ & $-2.65\substack{+0.63 \\ -0.83}$ & $1.50\substack{+0.21 \\ -0.42}$& $0.5$\\
        $1.50$ & $12.23\substack{+0.51 \\ -0.43}$ & $-3.12\substack{+0.25 \\ -0.26}$ & $1.50$& $0.5$ \\
        $2.50$ & $12.00\substack{+0.68 \\ -0.65}$ & $-3.11\substack{+0.45 \\ -0.35}$ & $1.50$& $0.5$ \\
        $4.05$ & $12.02\substack{+0.66 \\ -0.61}$ & $-3.67\substack{+0.44 \\ -0.35}$ & $1.50$& $0.5$ \\
        \hline
    \end{tabular}
\caption{Median-fit parameters for TIR LFs, $L^*$, $\phi^*$, $\alpha$, and $\sigma$ from equation \ref{eq:fit}, refer to Figure \ref{fig:TIR_LF_zbin_00}-\ref{fig:TIR_LF_zbin_03}. $\alpha$ is only fitted for $z=0-1$, and $\sigma$ is the fixed value from \protect\cite{2013MNRAS.432...23G}.}
\label{tab:2}
\end{table}


\subsection{Luminosity density evolution}\label{S:evolute}
The IR luminosity density, denoted by $\Omega$, depicts the average IR energy emitted by all galaxies within a specific age of the Universe. This is obtained by integrating the LF over the luminosity range, i.e.,
\begin{equation}
    \Omega = \int^{\infty}_{10^8} \phi (L) \, L \, d \log L
\end{equation}
based on the modified-Schechter function fit $\phi(L)$ for the LF. We integrate the LF down to $L^*=10^8 L_\odot$. This is the same as \cite{2013MNRAS.432...23G}, which has similar redshift coverage for comparison.
To derive the 1-sigma uncertainty of $\Omega$, we compute $\Omega$ for every individual model fit from the MCMC results (Figure \ref{fig:TIR_LF_zbin_00}-\ref{fig:TIR_LF_zbin_03}). Since a range of parameter sets are considered based on their probability distribution, we can ensure that the calculated distribution of $\Omega$ agrees with our LF fit. From this $\Omega$ distribution, the median, the 16th- and 84th percentile are obtained as the final 1-sigma uncertainty of $\Omega$. 
The horizontal error bar shows the redshift range of that bin, same as Figure \ref{fig:TIR_LF_evolution}.
The derived values of $\Omega$ for each redshift bin are given in Table \ref{tab:3}.

\begin{table}
    \centering
    \begin{tabular}{|c|c|c|c|}
        \hline
        $z$ & $\Omega$ & $\Omega_{7.7 \, \mu m}$ (All) & $\Omega_{7.7 \, \mu m}$ (SF-only)\\ 
            & & $\log\,[\Omega/L_\odot\,{\rm Mpc^{-3}}]$ & \\ \hline
        $0.50$ & $8.69\substack{+0.41 \\ -0.23}$ & $7.73\substack{+1.59 \\ -0.67}$ & $7.61\substack{+1.40 \\ -0.52}$ \\
        $1.50$ & $9.34\substack{+0.25 \\ -0.20}$ & $8.66\substack{+0.94 \\ -0.50}$ & $8.42\substack{+1.17 \\ -0.43}$ \\
        $2.50$ & $9.13\substack{+0.33 \\ -0.24}$ & $8.53\substack{+1.02 \\ -0.63}$ & $8.58\substack{+0.95 \\ -0.83}$ \\
        $4.05$ & $8.60\substack{+0.31 \\ -0.24}$ & $8.35\substack{+0.86 \\ -0.54}$ & - \\
        \hline
    \end{tabular}
\caption{IR and 7.7 $\mu$m luminosity density evolution, taken from Figure \ref{fig:TIR_density} and \ref{fig:7.7um_density}, respectively.}
\label{tab:3}
\end{table}

\begin{figure*}
    \centering
    \includegraphics[width=1.85\columnwidth]{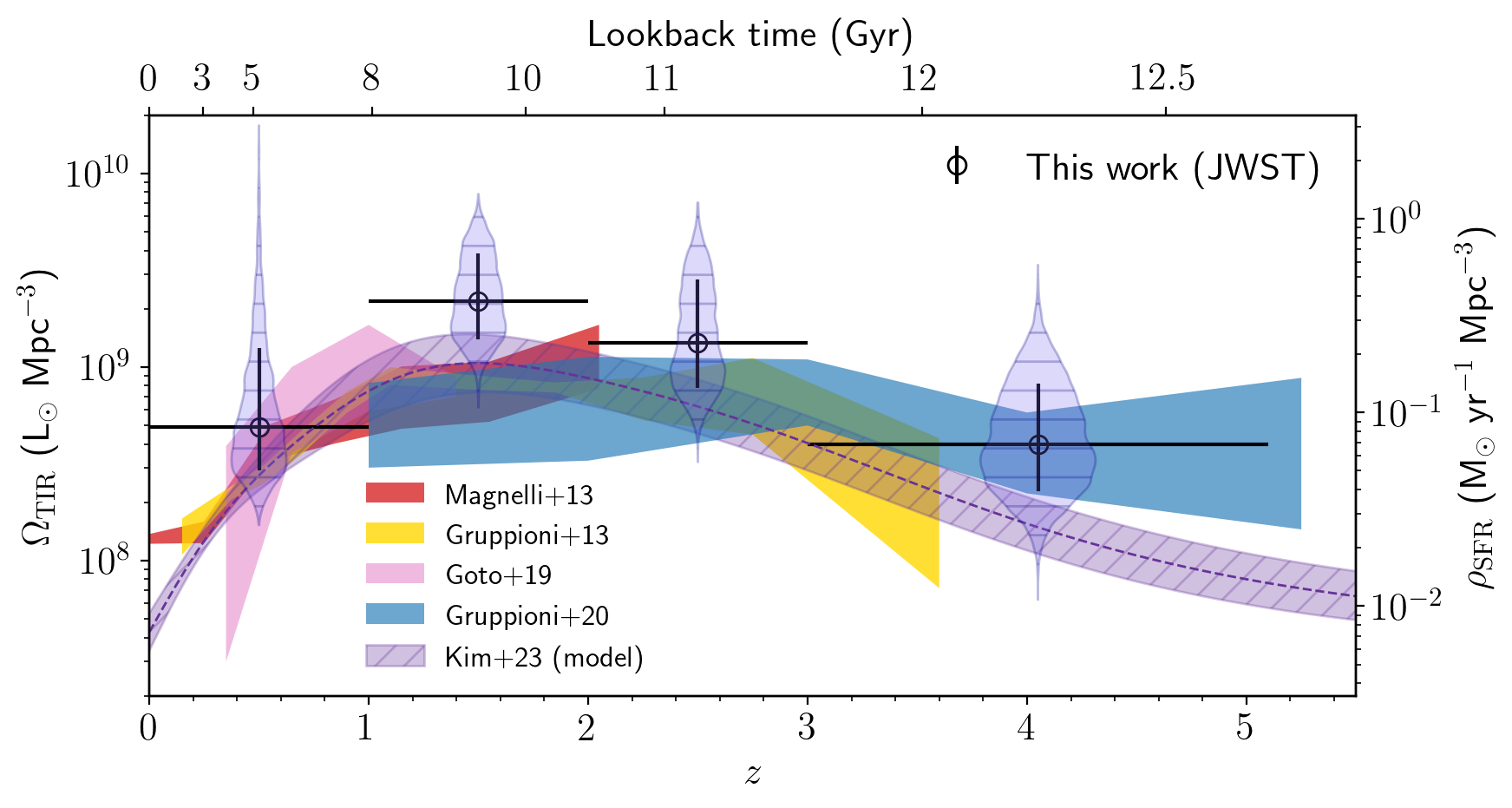}
    \caption{The redshift evolution of $\Omega$ (open circle). The purple shades are the violin plots that illustrate the probability distribution of $\Omega$. $\Omega$ evolution from the literature \protect\citep{2013A&A...553A.132M, 2013MNRAS.432...23G, 2019PASJ...71...30G, Gruppioni2020A&A...643A...8G, Kim2023} are also provided. SFR density ($\rho_{\rm SFR}$) converted from $\Omega$ is also shown for reference, assuming the relation from \protect\cite{Kennicutt1998ARA&A..36..189K}.}
    \label{fig:TIR_density}
\end{figure*}

Figure \ref{fig:TIR_density} presents the IR luminosity density ($\Omega$) evolution.
We obtained the probability distribution of $\Omega$ by integrating MCMC results.
For better visual interpretations, we overplot an associated "violin" (purple shades, centred at the bin) for each redshift bin, which displays the kernel density of $\Omega$ in a symmetric shape.
The width of the violin at a specific $\Omega$ value indicates the corresponding probability density.
We observe that the evolution of $\Omega$ well agrees with previous observational works \citep{2013A&A...553A.132M, 2013MNRAS.432...23G, 2019PASJ...71...30G, Gruppioni2020A&A...643A...8G} within their 1-sigma uncertainties, except for the $z=1-2$ bin. Our data points are generally higher than the model fit from \cite{Kim2023}, but the two error bars just touch each other.
The long tail distribution of $\Omega$ suggested by the violin plot apparently affects the estimation of its median and 1-sigma, because of the degeneracy in parameter fitting. If their mode, instead of median, is considered, the results would be more aligned with the literature.

We attributed the observed excess of $\Omega$ at $z=1-2$ to issues in the model fitting process that caused deviations in the fitted LF shape and consequently estimate of $\Omega$.
This is an example of fitting's inability to determine the break luminosity $L^*$ accurately. As Figure \ref{fig:TIR_LF_evolution} shows, an unusual bump in $L^*$ evolution can be identified at $z=1-2$.
Compared to \citet{2013A&A...553A.132M} and \citet{2013MNRAS.432...23G} which have been well constrained at this range, the bright end of our LF appears to be overestimated (refer to Figure \ref{fig:TIR_LF_zbin_01}).
Given our limited information on the bright end of the LF, $L^*$ is highly sensitive to the last luminosity bin that typically lies near the knee of our LF.
In the case of $z=1-2$, the last luminosity bin is relatively high with respect to the literature, potentially contributing to the observed phenomenon. 
We also note this effect is not significant for LF at other redshifts because their fitted bright end is generally consistent with the literature.

Our work pushes the MIR-based IR SF history to $z=4.05$ ($\sim 1.52$ Gyr after the Big Bang).
While IR SF history has already probed to $z\sim6$ with sub-mm data \citep{Gruppioni2020A&A...643A...8G}, it is still a significant improvement compared to the previous-generation IR space telescopes, where the highest redshifts of TIR LF they reached are $z\sim 1.4$ \citep[][\textit{AKARI}]{2019PASJ...71...30G} and $z\sim 3.6$ \citep[][\textit{Herschel}]{2013MNRAS.432...23G}.
Our results highlight the SFRD peak and the turnover at $z\sim 1.5$, and further reveal the dust-obscured SF history before the cosmic noon. 

We next investigate the luminosity density evolution in 7.7 $\mu$m due to its characteristic PAH emission. 
Although $L_{7.7\mu m}$ is expected to trace the star-formation rate and also $L_{\rm TIR}$, testing how far this correlation holds is of importance in understanding the relation between the far-IR dust emission and those from PAH molecules. Also, quantifying the density evolution of 7.7$\mu$m emission is crucial to plan future larger mid-IR surveys with \textit{JWST} and future telescopes such as the Far-IR Spectroscopy Space Telescope (FIRSST) and PRobe far-infrared mission for astrophysics \citep[PRIMA,][]{Moullet2023arXiv231020572M}.

\begin{figure}
    \centering
    \includegraphics[width=\columnwidth]{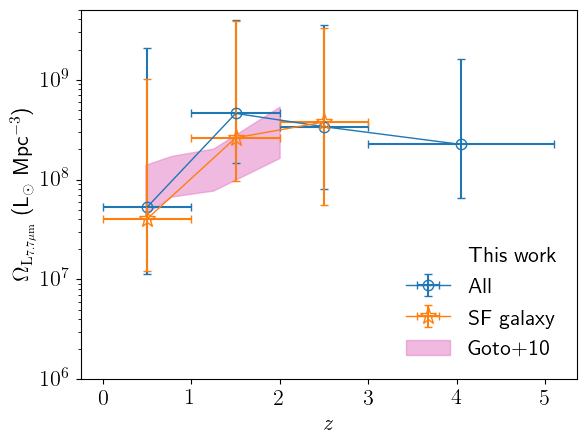}
    \caption{The redshift evolution of $\Omega_{7.7\, \mu m}$ for all galaxy (blue open circle) and only SF galaxy (orange open star). Comparison to \protect\cite{Goto2010} is provided.}
    \label{fig:7.7um_density}
\end{figure}

The density evolution of this specific luminosity has been previously studied by \cite{Goto2010}.
By fitting 7.7 $\mu$m LFs in Figure \ref{fig:mono_6LFs} and \ref{fig:mono_6LFs_SF} with the same method described in \S \ref{S:MCMC} and above, we compute the $\Omega_{7.7\, \mu m}$ for all galaxy and SF-only galaxy samples. 
The result is shown in Figure \ref{fig:7.7um_density}. 
First, we note our $\Omega_{7.7\, \mu m}$ evolution agrees with the trend of \cite{Goto2010} in their redshift range $z=0-2$. A peak, similar to $\Omega$, has been found at $z\sim1.5$.
Nevertheless, our $\Omega_{7.7\, \mu m}$ is severely affected by the insufficient data points in 7.7 $\mu$m LF that give huge degeneracy and uncertainty in the fitting. While the alignment of the $\Omega$ evolution of the two samples is expected as \S \ref{S:SF_AGN} discussed, no quantitative conclusions can be drawn on the differences or ratios between the two considering the overlapping error bars.
Since PAH emission has served as an important indicator of SF activity, studying PAH evolution and its contribution from SF / AGN galaxies is crucial for a comprehensive understanding of its behaviour and robustness across cosmic time. Here, we present a first step towards higher redshifts and encourage future works to investigate the PAH evolution with better statistics. 

It certainly requires a more extensive mid-IR survey to follow up on and confirm our results. We suggest utilising future mid-IR observations by \textit{JWST} with a larger survey area such as \cite{2017jwst.prop.1207R}. From a sample of thousands of galaxies, we expect to obtain the LF with higher resolution in redshift and reduced error.

\section{Conclusions}\label{S:conc}
In this work, we show the first \textit{JWST} rest-frame total IR LFs and MIR LFs at 7.7, 10, 12.8, 15, 18 and 21 $\mu$m at $z=0-5.1$ with MIR images from the CEERS survey.
While our LFs are consistent with previous studies on MIR LFs, we also present the LFs up to the highest redshift $z>4$ and to 2 orders of magnitude fainter in terms of IR luminosity.
For TIR LF, we found an overall density evolution of $\propto (1+z)^{-1.73\pm0.42}$. Due to the degeneracy in parameter fitting, the luminosity evolution shows a flatter shape of $\propto (1+z)^{0.90\pm0.98}$. The turnover of SFRD occurred at $z\sim1.5$.
With these results, we demonstrate the potential that \textit{JWST} can bring to future research on IR galaxy evolution. 

Still, we should exercise caution, as our LF may be undetermined at high-$z$ because MIRI only sampled the near-IR continuum, mainly contributed by stellar emission.
While scenarios that could reduce this effect have been discussed, supporting observations are surely needed for a more precise determination of \textit{JWST} galaxy IR LFs at high-$z$. 
Given that our LFs are based solely on photo-$z$ measurements, spec-$z$ from the \textit{JWST} NIRSpec and MIRI MRS will be crucial to minimise the redshift uncertainty in the calculation of IR luminosities and LFs for those faint \textit{JWST} galaxies. 

Since there is currently no far-IR space telescope with similar sensitivity to \textit{JWST}, counterpart observations in the sub-mm with ALMA bands will be essential to constrain the rest-frame far-IR SEDs. This is particularly effective for high-$z$ galaxies because the lowest luminosity bin of our LF at $z=3-5.1$ is already the same order of magnitude (by only $\sim$0.5 dex) as the LF reported by \cite{Gruppioni2020A&A...643A...8G} from ALMA data.
Furthermore, utilising the reddest MIRI band, F2550W (25.5 $\mu$m) in future surveys will also help to extend the redshift range for probing rest-frame MIR emission. 
These findings motivate further MIRI and corresponding sub-mm observations in order to deepen our view of the obscured star-forming history of the Universe.

\section*{Acknowledgements}
The authors are grateful to the anonymous referee for the valuable and constructive comments, which significantly improved the paper. 
The authors appreciate the suggestion from Denis Burgarella and Guang Yang for improving the result of SED fitting.
The authors would like to thank Shotaro Yamasaki and Toshifumi Futamase for their suggestions and comments. 
TG acknowledges the support of the National Science and Technology Council of Taiwan through grants 108-2628-M-007-004-MY3, 111-2112-M-007-021, 112-2112-M-007-013, and 112-2123-M-001-004-.
TH acknowledges the support of the National Science and Technology Council of Taiwan through grants 110-2112-M-005-013-MY3, 110-2112-M-007-034-, and 112-2123-M-001-004-.
SH acknowledges the support of The Australian Research Council Centre of Excellence for Gravitational Wave Discovery (OzGrav) and the Australian Research Council Centre of Excellence for All Sky Astrophysics in 3 Dimensions (ASTRO 3D), through project number CE17010000 and CE170100013, respectively. 
This work is based on observations made with the NASA/ESA/CSA James Webb Space Telescope. The data were obtained from the Mikulski Archive for Space Telescopes at the Space Telescope Science Institute, which is operated by the Association of Universities for Research in Astronomy, Inc., under NASA contract NAS 5-03127 for \textit{JWST}. These observations are associated with program JWST-ERS01345.
The authors acknowledge the CEERS team for developing their observing program with a zero-exclusive-access period.
This work is based on observations taken by the CANDELS Multi-Cycle Treasury Program with the NASA/ESA \textit{HST}, which is operated by the Association of Universities for Research in Astronomy, Inc., under NASA contract NAS 5-26555.
This work used high-performance computing facilities operated by the Center for Informatics and Computation in Astronomy (CICA) at National Tsing Hua University. This equipment was funded by the Ministry of Education of Taiwan, the National Science and Technology Council of Taiwan, and National Tsing Hua University.

\section*{Data Availability}
The MIRI observations from \textit{JWST} CEERS survey are publicly available at the MAST archive \url{https://mast.stsci.edu/portal/Mashup/Clients/Mast/Portal.html}.
The EGS Multi-Band Source and Photometric Redshift catalogue can be downloaded at \url{https://archive.stsci.edu/hlsp/candels/egs-catalogs}.
Other data underlying this article will be shared upon reasonable request to the corresponding author.


\bibliographystyle{mnras}
\bibliography{jwstLF} 



\appendix
\section{Example of fitted SEDs}
We present examples of fitted SED from \textsc{cigale} in the following Figure \ref{fig:sed_example_00}-\ref{fig:sed_example_03} for readers to verify the result. In each redshift bin, we randomly pick up SEDs for one SF galaxy and one AGN host galaxy based on frac$_{\rm AGN}$, as explained in \S \ref{S:sed}. The wavelength coverage of plotted SEDs is $0.3-6000$ $\mu$m, i.e., covering the whole IR range. 

\begin{figure*}
    \centering
    \includegraphics[width=1.75\columnwidth]{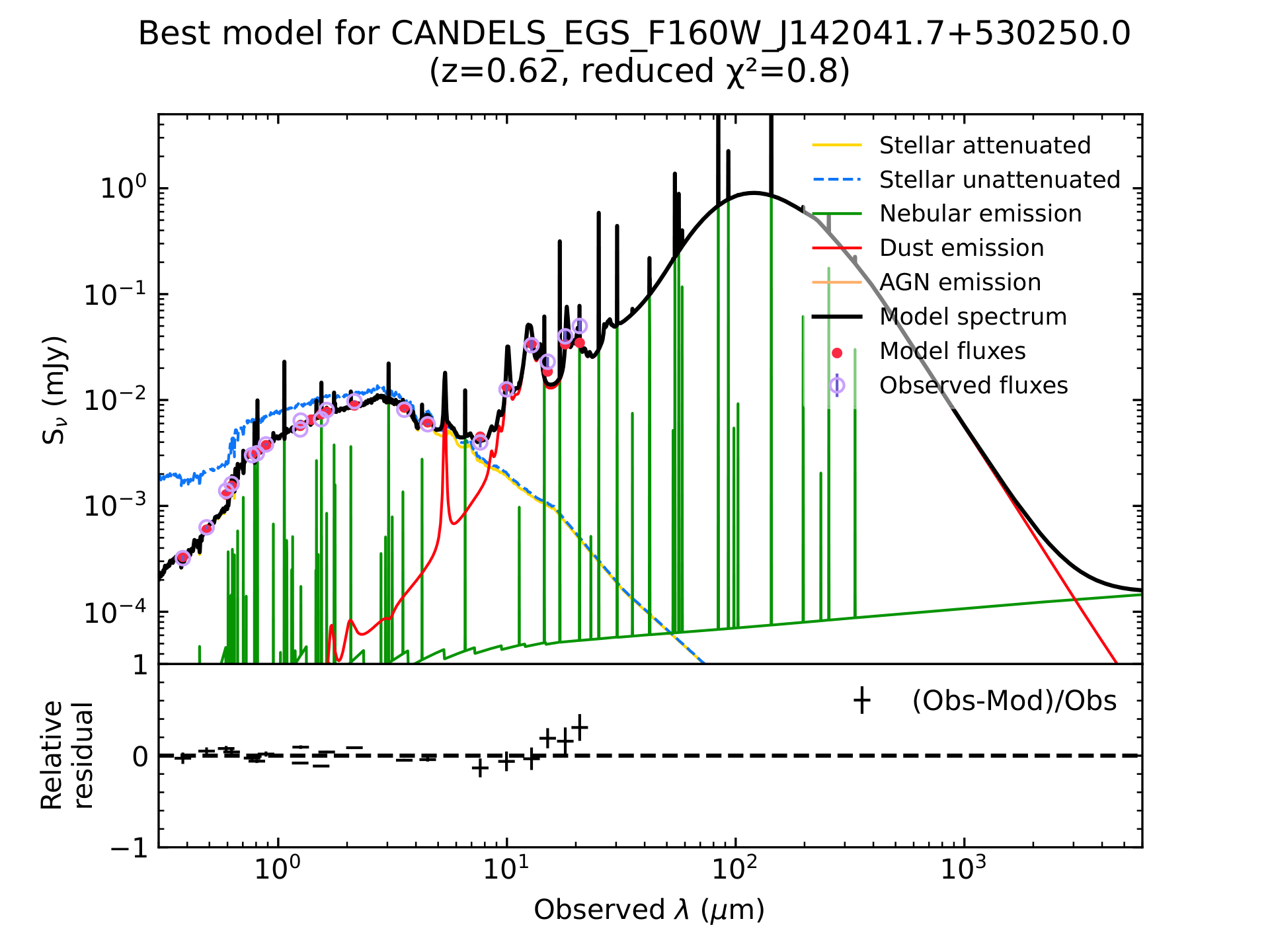}
    \includegraphics[width=1.75\columnwidth]{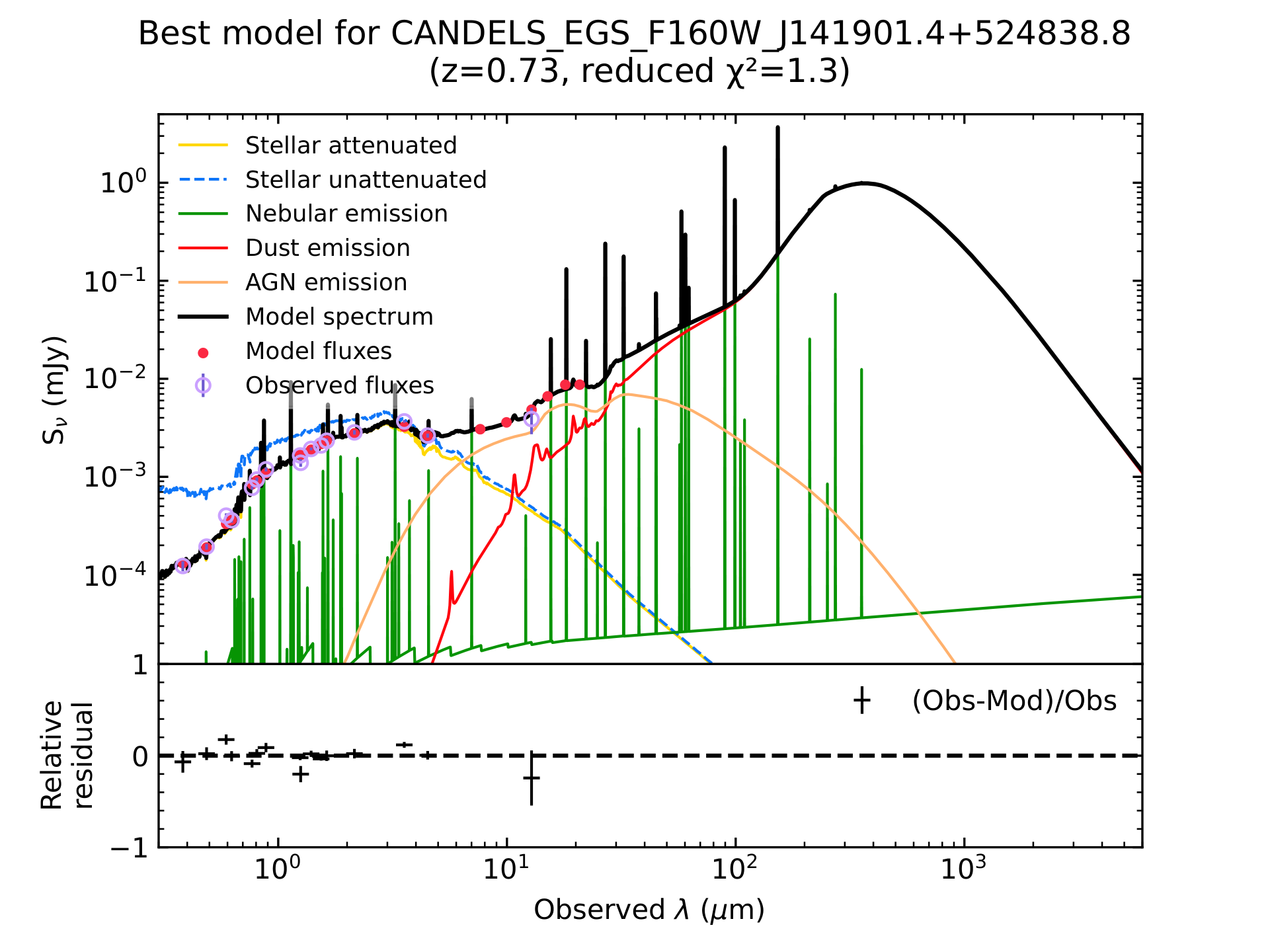}
    \caption{Example fitted galaxy SEDs (in observed frame) from \textsc{cigale} for $z=0-1$ bin, with their photo-$z$ and reduced $\chi^2$. The left panel shows a star-forming galaxy and the right panel shows an AGN host galaxy. The IAU designation from the CANDELS-EGS catalogue is provided as id.}
    \label{fig:sed_example_00}
\end{figure*}

\begin{figure*}
    \centering
    \includegraphics[width=1.75\columnwidth]{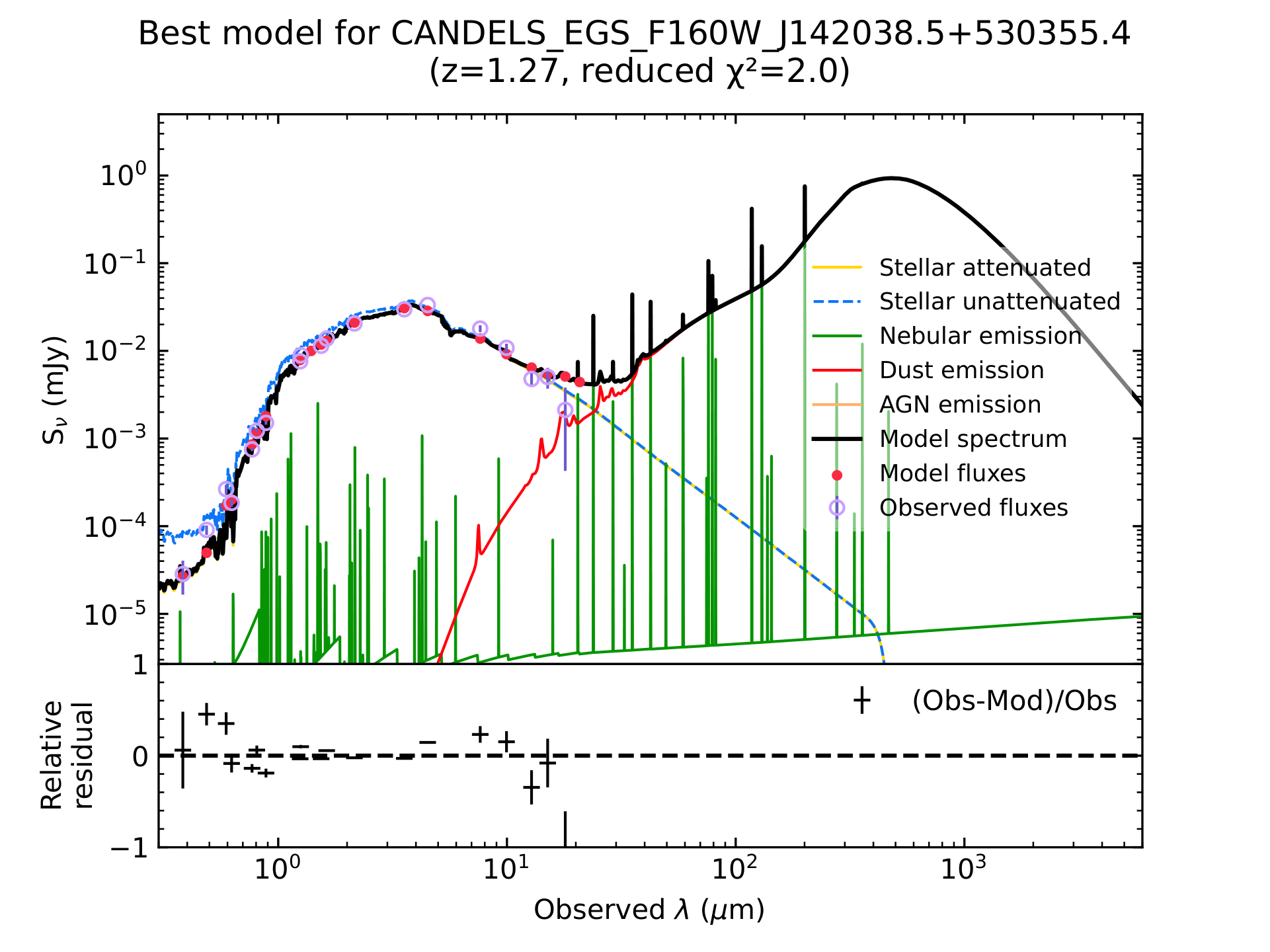}
    \includegraphics[width=1.75\columnwidth]{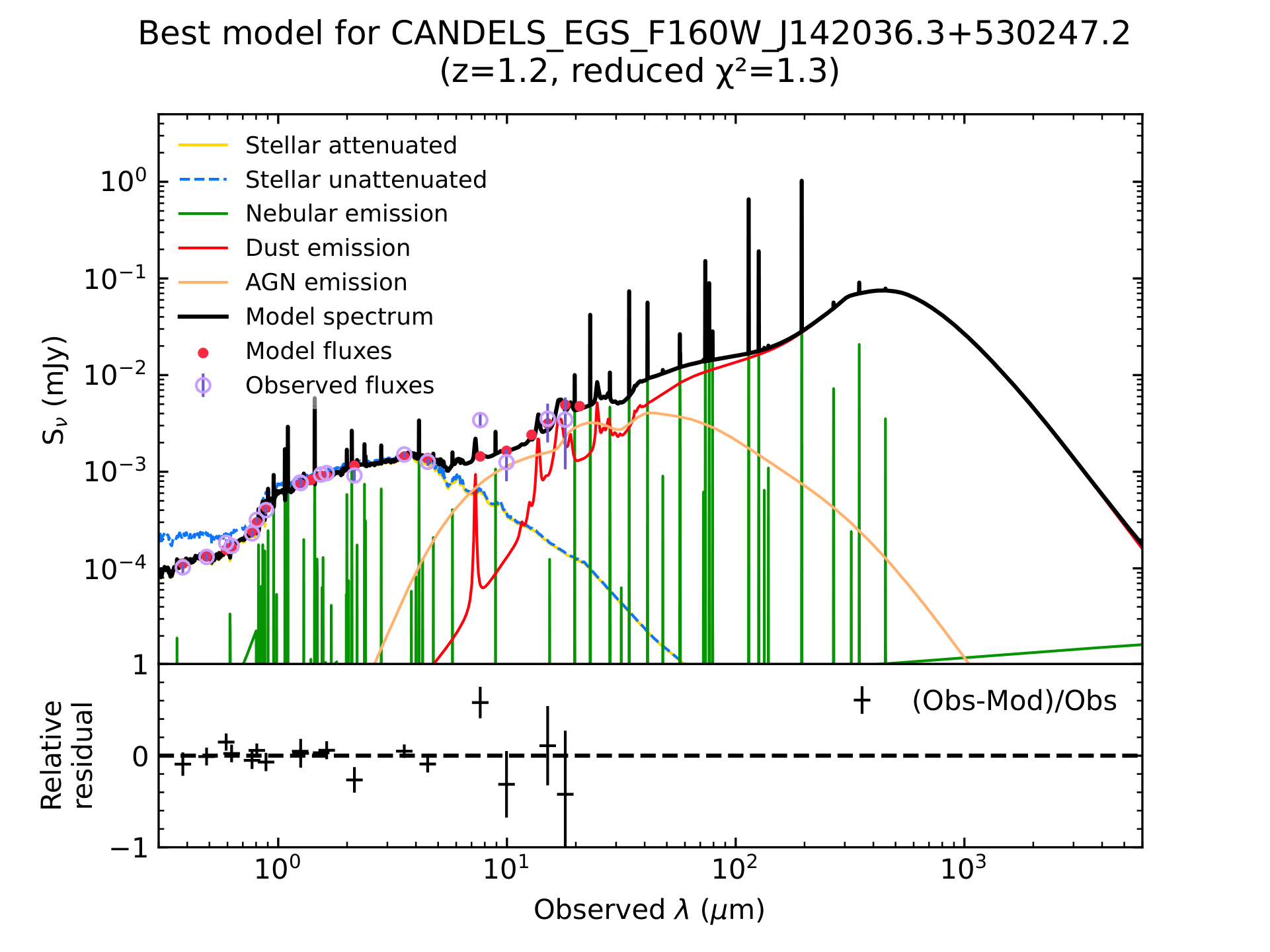}
    \caption{Same as Figure \ref{fig:sed_example_00} but for $z=1-2$ bin.}
    \label{fig:sed_example_01}
\end{figure*}

\begin{figure*}
    \centering
    \includegraphics[width=1.75\columnwidth]{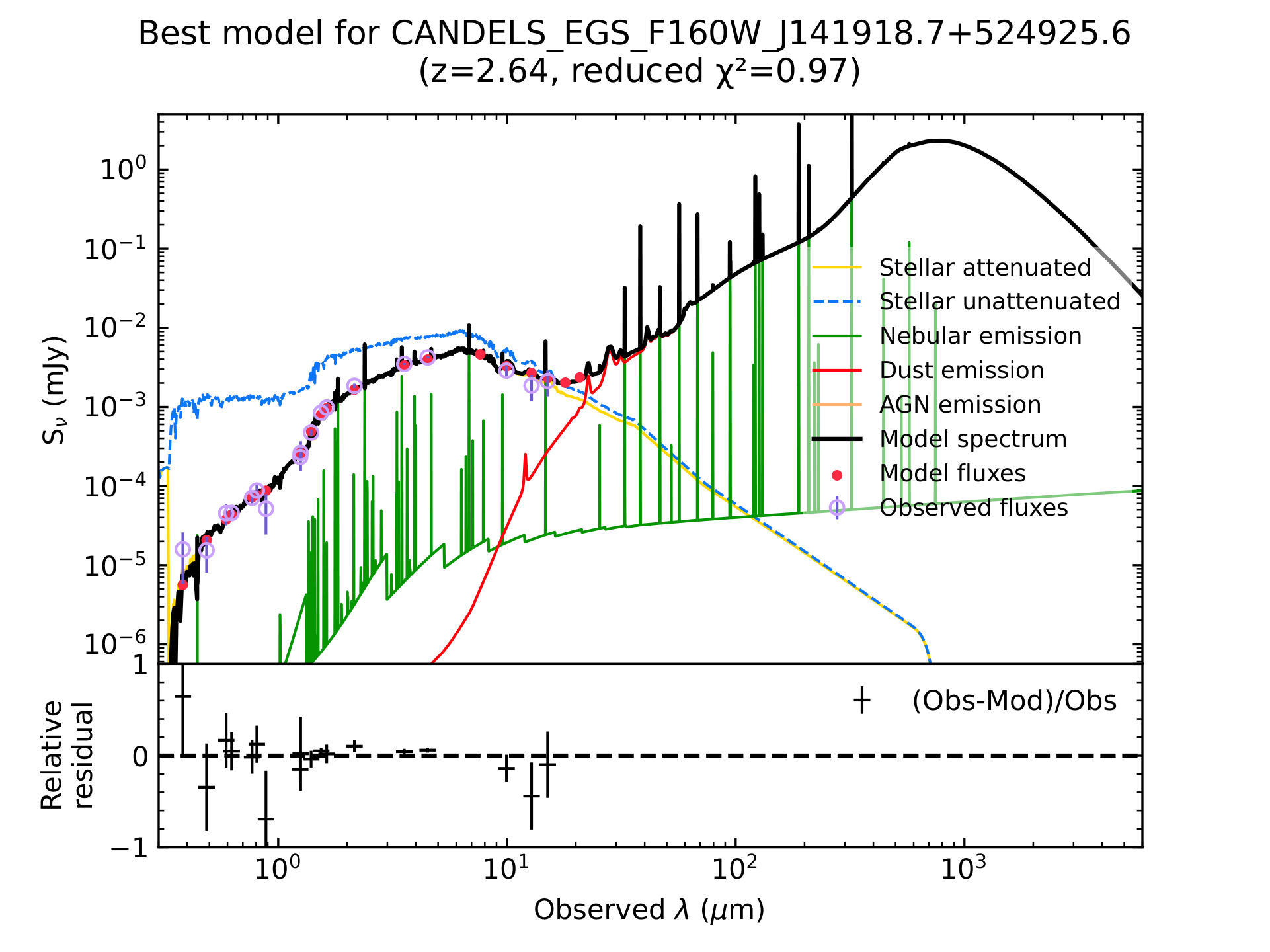}
    \includegraphics[width=1.75\columnwidth]{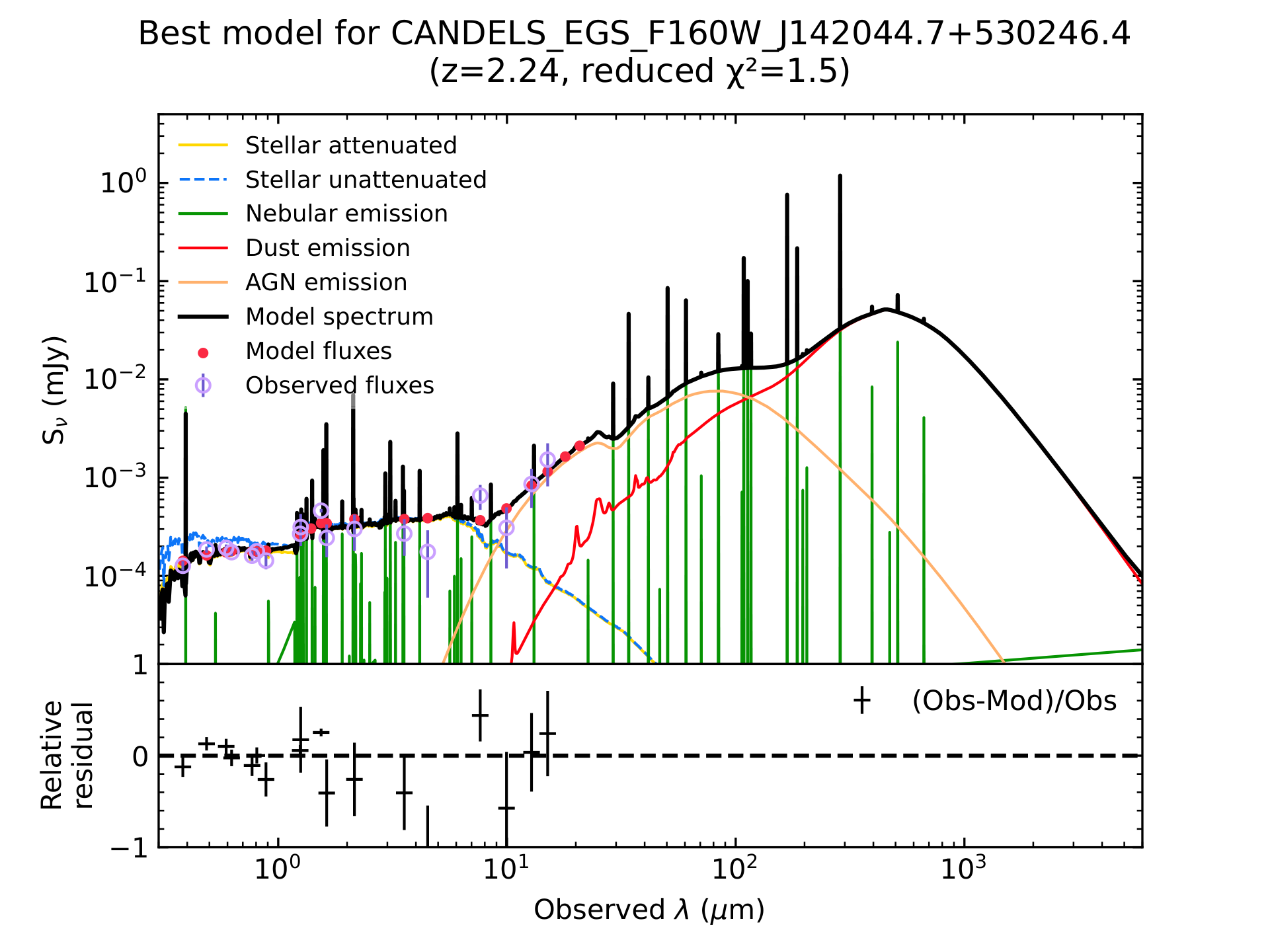}
    \caption{Same as Figure \ref{fig:sed_example_00} but for $z=2-3$ bin.}
    \label{fig:sed_example_02}
\end{figure*}

\begin{figure*}
    \centering
    \includegraphics[width=1.75\columnwidth]{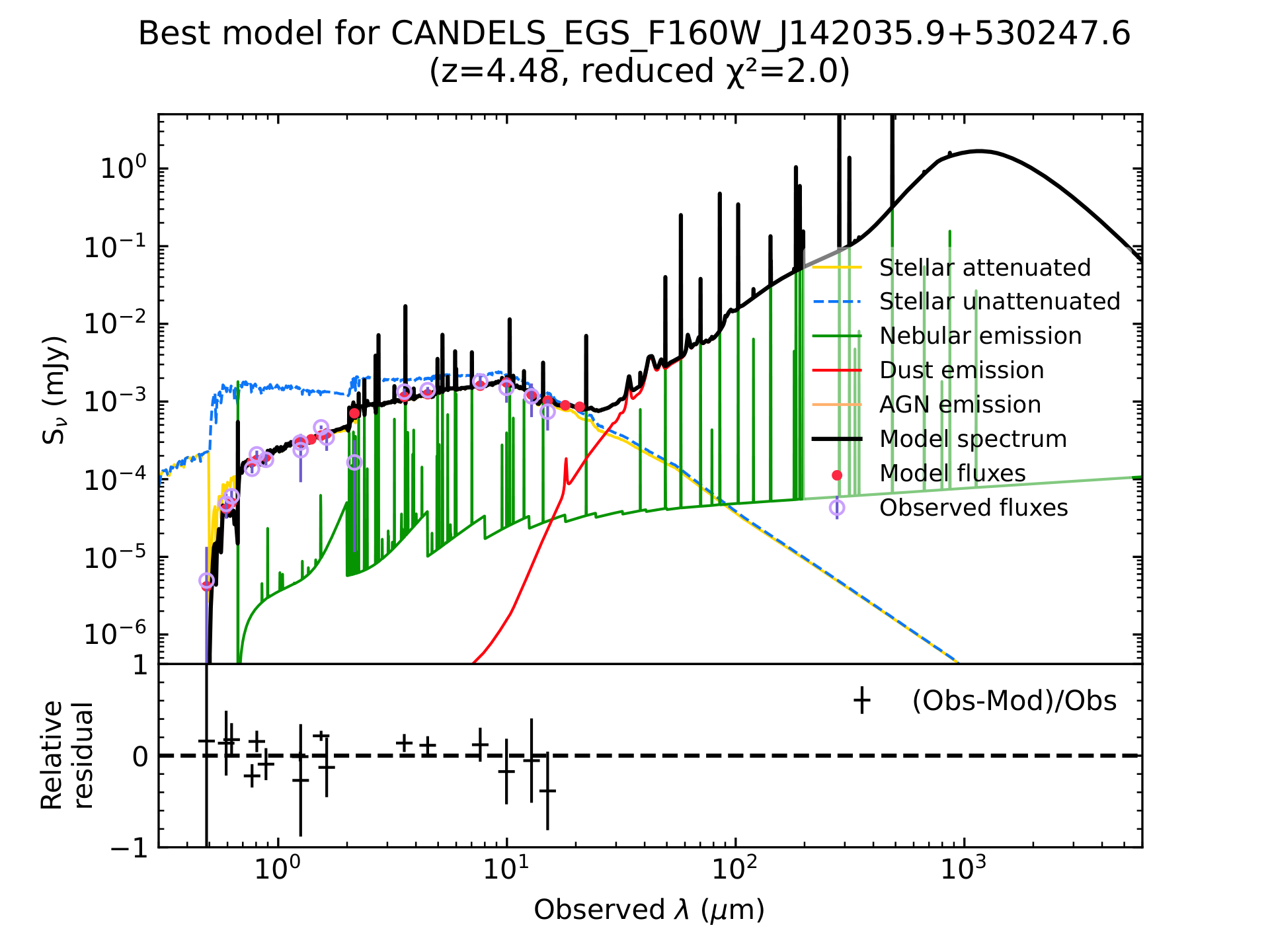}
    \includegraphics[width=1.75\columnwidth]{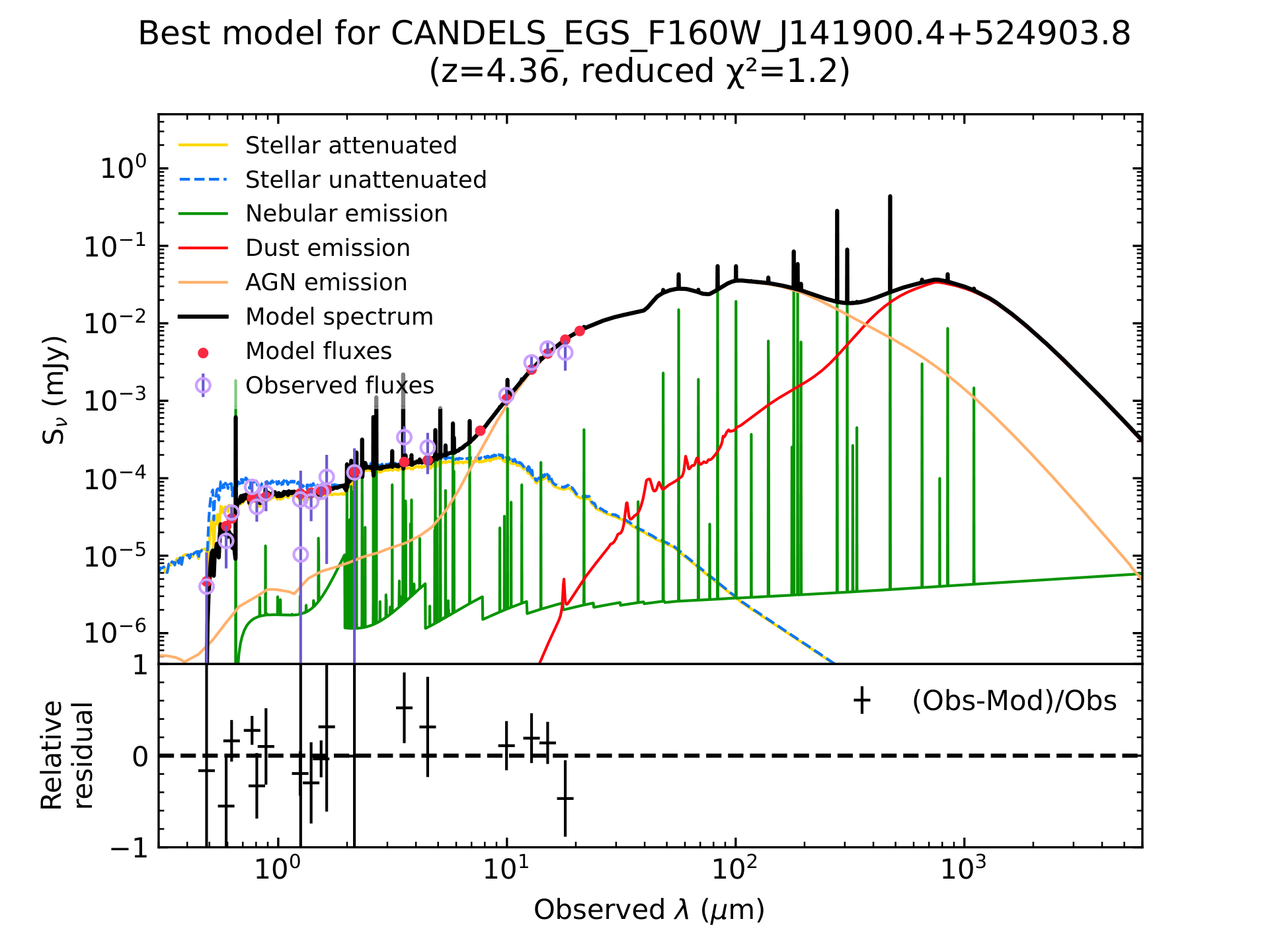}
    \caption{Same as Figure \ref{fig:sed_example_00} but for $z=3-5.1$ bin.}
    \label{fig:sed_example_03}
\end{figure*}



\bsp	
\label{lastpage}
\end{document}